\documentclass[aip,jcp,nobibnotes,twocolumn,floatfix,reprint]{revtex4-1}%
\usepackage{amsmath}
\usepackage{graphicx}
\usepackage{amsfonts}
\usepackage{amssymb}
\usepackage{color}
\usepackage{comment}%
\providecommand{\U}[1]{\protect\rule{.1in}{.1in}}
\begin{document}
\title{On the dual structure of the Schr\"{o}dinger dynamics}
\author{Kazuo Takatsuka}
\thanks{kaztak@fukui.kyoto-u.ac.jp}
\affiliation{Fukui Institute for Fundamental Chemistry, Kyoto University, 606-8103 Kyoto, Japan.}

\begin{abstract}
This paper elucidates the dual structure of the Schr\"{o}dinger dynamics in
two correlated stages: (1) Derivation of the real-valued Schr\"{o}dinger
equation from scratch in a universal manner without referring to classical
mechanics, wave mechanics, nor optics. Also, a real-valued path-integral is
formulated as the Green function of the real-valued Schr\"{o}dinger equation,
which bears the Wiener measure in contrast to the Feynman path integrals.
These derivations reveal the clear and universal interpretation of the
Schr\"{o}dinger equation and function. (2) We then study a quantum stochastic
path dynamics in a manner compatible with the Schr\"{o}dinger equation. The
relation between them is like the Langevin dynamics and the diffusion
equation. Each quantum path describes a \textquotedblleft
trajectory\textquotedblright\ in configuration space, representing, for
instance, a singly launched electron in the double-slit experiment that leaves
a spot one by one at the measurement board. Meanwhile the accumulated spots
give rise to the fringe pattern as predicted by the absolute square of the
Schr\"{o}dinger function. In an analogy to stochastic dynamics, we thus see
the global picture of quantum dynamics in a triangle relation between the
real-valued Schr\"{o}dinger equation as a pair of coupled diffusion equations,
the real-valued path integrals in a matrix form as the quantum mechanical
extension of the Feynman-Kac formula, and the quantum stochastic path dynamics
represented in the Ito stochastic differential equation. The physical
significance of the quantum stochasticity such as the origin of the
uncertainty relations, the indirect correlation among the quantum paths,
quantum entanglement and detanglement, and so on, are discussed. The
self-referential nonlinear interrelationship between the Schr\"{o}dinger
functions (regarded as a whole) and the quantum paths (as its parts) is
identified as the ultimate mystery in quantum dynamics.

\end{abstract}
\date{\today}
\maketitle


\color{blue}\color{black}

\section{Introduction\label{sec:Introduction}}

Even one hundred years after the birth, understanding and interpretation about
quantum mechanics\cite{dirac1981principles,Schiff,Bohm-text,
Messiah,schwinger2001quantum,Feynman-Hibbs,ruetsche2011interpreting,omnes1992consistent}
remain unsettled, with no tight consensus to this
day.\cite{auletta2001foundations,home2007einstein,whitaker2012new,freire2022oxford}
The controversy often centers around the mysterious nature of the
Schr\"{o}dinger (wave) function. Besides, it is a common practice to rest on
the Schr\"{o}dinger equation alone in the study of any of quantum dynamics.

In reality, however, the Schr\"{o}dinger dynamics has a dual structure; the
dynamics described by the Schr\"{o}dinger equation and the dynamics subjected
to quantum path dynamics. The critical role of the dual structure is well
illustrated in terms of the double-slit experiment. It is well known that the
repeated launch of electrons one-by-one leaves the so-called interference
fringe patterns in the spot distribution made by very many electrons. The
pattern is well reproduced and predicted by the absolute square of the
Schr\"{o}dinger functions. Yet, the Schr\"{o}dinger equation does not
necessarily care about the dynamics of individual particles, and it tells
virtually nothing about the relevant on-going process. Therefore, the double
slit experiment bears two mysteries in it; one is why the Schr\"{o}dinger
function behaves as though it is subjected to the Huygens principle (note that
it is never obvious that the linearity of the Schr\"{o}dinger function
warrants this property).\cite{Goussev-Huygens,takatsuka2023schrodinger} The
other one is why and how the one-by-one ejection of electrons reproduces the
fringe patter after all. These are correlated but are different matters. \color{black}

In diffusion theory, we first have the diffusion differential equation
describing the probability distribution function of the relevant particle
position, which is followed by the Wiener path integrals and/or the
Feynman-Kac formula as the Green function of the diffusion differential
equations. Besides, the Langevin dynamics and Ito stochastic differential
equation describe the path dynamics of an individual Brownian particle. These
constitute a triangle relationship for the representation of entire stochastic
dynamics. In turn to quantum dynamics, we have the Schr\"{o}dinger
differential equation and the Feynman path integrals, which are both complex
valued. Yet, the quantum path dynamics is generally disregarded. (This should
not be confused with the semiclassical path
representation\cite{Messiah,Schulman,U-Miller1,Smilansky,
kleinert2006path,Maslov,Brack,Paper-I,Paper-II} or the Bohmian trajectory
representation of the Schr\"{o}dinger
function.\cite{bohm1952suggested,Bohm-text,Wyatttext,sanz2013trajectory}) To
comprehend the entire Schr\"{o}dinger dynamics, we need to find out the
similar triangle structure and the properties arising from it.

In light of the mathematical similarity of the Schr\"{o}dinger equation to the
diffusion equation, we may expect the similar theoretical framework in the
quantum dynamics: (I-1) The real-valued Schr\"{o}dinger equation as a coupled
diffusion equations to determine the relevant distributions (the density
function and the Schr\"{o}dinger (wave) function), (I-2) A real-valued path
integrals in the matrix form as the Green function to the real-valued
Schr\"{o}dinger equation, which suggest the presence and the role of the
stochastic paths behind the distribution function. And (II) the quantum
stochastic path dynamics in the form of Ito stochastic differential equation,
the velocity drift term of which is composed of the local velocity field
attained in the real-valued Schr\"{o}dinger dynamics. We thus try to fill in
the gaps in the framework of Schr\"{o}dinger dynamics that cannot be fully
understood through the Schr\"{o}dinger equation alone.

The precise issues addressed to achieve the goal are as follows. (1)
Derivation of the real-valued Schr\"{o}dinger equation from scratch. Although
the Schr\"{o}dinger equation is already well established, we would like to
restart from the very bottom in order to link to the quantum path dynamics.
(2) We propose a path integral representation as the Green function for the
real-valued Schr\"{o}dinger equation. (3) Then follows the derivation of
quantum stochastic path dynamics based on the Ito stochastic differential
equation and the Feynman-Kac formula of statistical mechanics. (4) Finally,
outcomes from the self-referential nonlinear relations between the
Schr\"{o}dinger function and quantum stochastic paths will be discussed.

The broader aspects of the above three items are as follows.

(1) To begin with, the presence of the imaginary number unit $i=\sqrt{-1}$ in
the Schr\"{o}dinger equation seems to make its physical meaning and foundation
rather obscure and biased. In fact, the imaginary numbers do not appear in the
quantum stochastic path dynamics. We therefore rebuild the real-valued
Schr\"{o}dinger equation in a manner free from any of the historical
interpretations.\cite{auletta2001foundations,home2007einstein,whitaker2012new,freire2022oxford}
Particularly, we deny the idea that the Schr\"{o}dinger functions represent
some kind of \textquotedblleft wave\textquotedblright, since the notion of
wave\ will not lead to that of quantum stochastic paths. Moreover, we should
seek for a universal way to reach the Schr\"{o}dinger equation, simply because
the Schr\"{o}dinger equation itself should be more fundamental than others. We
start from the vector factorization of the real-space distribution function
and impose the minimal physical requirements to attain the Schr\"{o}dinger
equation. The \textquotedblleft local velocity field\textquotedblright%
\ obtained in this process turns out to corresponds to the velocity drift term
in the Ito stochastic differential equation for quantum dynamics.

(2) A path-integral in real configuration space is a quantum mechanical
extension of the Feynman-Kac formula. It has the Wiener measure, and thereby
clarifies how quantum mechanics and stochastic one are theoretically similar
to each other. The present path integrals naturally guide us to an idea that
quantum dynamics should be supported by a set of stochastic paths.

(3) Our derivation of quantum stochastic path dynamics follows the pioneering
work by Nagasawa,\cite{nagasawa1989transformations,Nagasawa2} who established
a quantum theory based on the one-to-one correspondence between parabolic
real-valued partial differential equations and the Ito stochastic equations.
Here in this work, we start from the relationship between the Feynman-Kac
formula, \cite{kac1949distributions,del2004feynman,Ezawa} Ito stochastic
differential equation, and the Schr\"{o}dinger equation. The attained
stochastic differential equation consists of the velocity drift term and the
Wiener process having \textquotedblleft quantum diffusion
constant\textquotedblright. It turns out that the velocity drift terms appears
to be a function of the Schr\"{o}dinger function.

(4) The total ensemble of thus generated quantum stochastic paths should
reproduce the overall dynamics of the Schr\"{o}dinger function, and therefore
the individual quantum paths and the Schr\"{o}dinger function are in a
self-referential nonlinear relationship. There is no coherent interaction
among the quantum paths. Yet, through the nonlinear relationship the
individual quantum paths are indirectly correlated with each other, explaining
why one-by-one electron launching in the double-slit experiment leads to the
fringe pattern after accumulation of all.

(5) Based on the above findings we discuss the physical consequences of the
stochasticity in quantum dynamics. First we clarify that the
quantum-mechanical Wiener process serves as an intrinsic element to quantum
path dynamics. To illustrate it, we show that the scaling law in the Wiener
process reproduces the energy eigenvalues of the hydrogen atom and the Bohr
radius with no use of the Bohr model or the de Broglie wavelength. The
stochasticity is also regarded as one of the physical origins of the
uncertainty relation. In an effort to clarify the physical meaning of the
velocity drift term, we derive the quantum-mechanical Hamilton canonical
equations of motion and the Newtonian equation for quantum stochastic path
dynamics, highlighting the fundamental differences between the quantum paths
and classical trajectories. With respect to the quantum entanglement and
associated symmetry, we suggest that the Wiener process can break the
entanglement (detangle) even in case where the Schr\"{o}dinger equation has no
mechanism to do so.

This paper does not intend to propose a new interpretation about quantum
mechanics nor treat cutting-edge quantum technology. Rather it intends to
clarify the basic structure of the Schr\"{o}dinger dynamics, which should be
known before those studies are to be made. This work is structured as follows:
We derive the real-valued Schr\"{o}dinger equation and study the relevant
properties in Sec. \ref{sec:RealValue} and the real-valued quantum path
integrals in Sec. \ref{sec:Feynman-Kac}. Section \ref{sec:Stochastic} follows
with the quantum stochastic path dynamics. Some significant properties arising
from the quantum stochasticity are shown in Sec. \ref{Sec. Wiener}. The
properties of the quantum stochastic paths are featured in Sec.
\ref{sec:correlation}. The present paper concludes with some remarks in Sec.
\ref{sec:Conc}.

\section{Real-valued Schr\"{o}dinger equation from scratch and the
Schr\"{o}dinger vectors as a coherent distribution
function\label{sec:RealValue}}

It is well known that Schr\"{o}dinger figured out his equation rather
intuitively without a concrete theoretical
ground.\cite{auletta2001foundations,home2007einstein,whitaker2012new,freire2022oxford}
Its background relation to the Hamilton-Jacobi equation and a similarity to
the geometrical optics have been widely pointed
out.\cite{simeonov2024derivation} An interesting twist, which is relevant to
the present paper, is due to Nelson's quantization based on his proposed the
stochastic Newtonian equation.\cite{Nelson,nelson2012review} There have been
proposed other theories backing the Schr\"{o}dinger equation, each of which
gives its own physical and/or mathematical interpretations about the
Schr\"{o}dinger equation.\cite{auletta2001foundations,freire2022oxford} Yet,
the most commonly accepted idea to the Schr\"{o}dinger equation thus far is to
regard it as an axiom or the undisputed foundation of physics in the quantum
scales. Nevertheless, we here propose a universal way of derivation of the
Schr\"{o}dinger equation\cite{takatsuka2025analysis} and the associated path
integrals in the real field, one of the purposes of which is to make a smooth
continuation to the quantum path dynamics. \color{black}

\subsection{Real-valued Schr\"{o}dinger equation}

\subsubsection{Density distribution function\label{Sec:ConfigSpace}}

Suppose we have $\rho(q,t)$, a general density distribution function in
configuration space $q$ and time $t$, which is positive semidefinite
everywhere and can be normalized such that
\begin{equation}
\int dq\rho(q,t)=1.
\end{equation}
We impose the following basic and universal constraints on $\rho(q,t)$.

1) The translational invariance of configuration space $q$, which is assumed
to be Euclidean.

2) Translational invariance of time.

3) Incompressibility of $\rho(q,t)$.

Notice that all these conditions and the definition of $\rho(q,t)$ are general
and not peculiar to quantum mechanics.

\subsubsection{Factorization of $\rho(q,t)$ and its dynamics
\label{subsec:factoraliztion}}

Since $\rho(q,t)\geq0$ everywhere, one can factorize it as%

\begin{equation}
\rho(q,t)=\left(
\begin{array}
[c]{cc}%
\phi_{r}(q,t) & \phi_{c}(q,t)
\end{array}
\right)  \left(
\begin{array}
[c]{c}%
\phi_{r}(q,t)\\
\phi_{c}(q,t)
\end{array}
\right)  =\bar{\psi}(q,t)^{T}\bar{\psi}(q,t) \label{fact}%
\end{equation}
with%
\begin{equation}
\bar{\psi}(q,t)\equiv\left(
\begin{array}
[c]{c}%
\phi_{r}(q,t)\\
\phi_{c}(q,t)
\end{array}
\right)  . \label{fact2}%
\end{equation}
We may choose both $\phi_{r}(q,t)$ and $\phi_{c}(q,t)$ to be real-valued
functions. We refer to $\bar{\psi}(q,t)$ as the Schr\"{o}dinger vector. The
factorizing vectors $\bar{\psi}(q,t)$ can have higher components than two.
However, it turns out later that the two component Schr\"{o}dinger vectors are
large enough to reproduce (nonrelativistic) quantum dynamics. We below examine
the effects of the above three conditions on the factorization.

\subsubsection{Translational invariance of free configuration space}

Shift the coordinate $q\rightarrow q+\Delta q$,$\,\ $and $\bar{\psi}(q,t)$ is
translated as%
\begin{equation}
\bar{\psi}(q,t)\rightarrow\tilde{\psi}(q+\Delta q,t).
\end{equation}
Due to the inherent freedom of the vector rotation of $\bar{\psi}(q,t)$, it
should generally undergo the following transformation on the occasion of the
$q$-shift%

\begin{align}
\tilde{\psi}(q+\Delta q,t)  &  =\exp\left(  \Delta q\text{\textbf{A}}%
\vec{\nabla}_{q}\right)  \bar{\psi}(q,t)\nonumber\\
&  =\left(  \text{\textbf{I}}+\Delta q\text{\textbf{A}}\vec{\nabla}%
_{q}\right)  \left(
\begin{array}
[c]{c}%
\phi_{r}(q,t)\\
\phi_{c}(q,t)
\end{array}
\right)  +\left(  \text{higher}\right)  , \label{q-shift}%
\end{align}
where \textbf{A} is a $2\times2$ real matrix, and%
\begin{equation}
\tilde{\psi}(q+\Delta q,t)^{T}=\left(
\begin{array}
[c]{cc}%
\phi_{r}(q,t) & \phi_{c}(q,t)
\end{array}
\right)  \exp\left(  \Delta q\text{\textbf{A}}^{T}\vec{\nabla}_{q}\right)  ,
\end{equation}
with $T$ indicates the matrix transposition. In Eq. (\ref{q-shift}),
\textbf{A}$\vec{\nabla}_{q}$ serves as the displacement operator matrix. The invariance%

\begin{equation}
\tilde{\psi}(q+\Delta q,t)^{T}\tilde{\psi}(q+\Delta q,t)=\bar{\psi}%
(q,t)^{T}\bar{\psi}(q,t)
\end{equation}
demands that%

\begin{align}
&  \tilde{\psi}(q+\Delta q,t)^{T}\tilde{\psi}(q+\Delta q,t)-\bar{\psi
}(q,t)^{T}\bar{\psi}(q,t)\nonumber\\
&  =\left(
\begin{array}
[c]{cc}%
\phi_{r}(q,t) & \phi_{c}(q,t)
\end{array}
\right)  \left[  \text{\textbf{A}}^{T}\vec{\nabla}_{q}+\text{\textbf{A}}%
\vec{\nabla}_{q}\right]  \left(
\begin{array}
[c]{c}%
\phi_{r}(q,t)\\
\phi_{c}(q,t)
\end{array}
\right)  \Delta q=0, \label{q-trans}%
\end{align}
which requires%

\begin{equation}
\text{\textbf{A}}^{T}=-\text{\textbf{A.}}%
\end{equation}
We may thus set%
\begin{equation}
\text{\textbf{A}}=c_{p}\text{\textbf{J,}}%
\end{equation}
where \textbf{J} is a 2$\times$2 unit symplectic matrix (or called the
standard symplectic matrix)\cite{Arnold,de2006symplectic} defined as%
\begin{equation}
\text{\textbf{J}}\mathbf{=}\left(
\begin{array}
[c]{cc}%
0 & -1\\
1 & 0
\end{array}
\right)  .
\end{equation}

The spirit\ of Noether theorem\cite{Arnold} and the discussion about the
linear momentum in terms of the displacement operator by
Dirac\cite{dirac1981principles} suggest that we may define the momentum
operator matrix (simply momentum operator) as%

\begin{equation}
\hat{p}=c_{p}\text{\textbf{J}}\vec{\nabla}_{q}. \label{p-op}%
\end{equation}
Equation (\ref{q-trans}) suggests that $\hat{p}$ should be placed in such a
position that%
\begin{equation}
\left(
\begin{array}
[c]{cc}%
\phi_{r}(q,t) & \phi_{c}(q,t)
\end{array}
\right)  \hat{p}\left(
\begin{array}
[c]{c}%
\phi_{r}(q,t)\\
\phi_{c}(q,t)
\end{array}
\right)  . \label{p-op2}%
\end{equation}

\subsubsection{Translational invariance in time\label{subsec:variation}}

The similar translational invariance is applied to the time coordinate such that%

\begin{equation}
\tilde{\psi}(q,t+\Delta t)^{T}\tilde{\psi}(q,t+\Delta t)=\bar{\psi}%
(q,t)^{T}\bar{\psi}(q,t).
\end{equation}
As above, the time-displacement operator is tied to the energy operator
$\hat{H}$ in the form%
\begin{equation}
\hat{H}=c_{t}\text{\textbf{J}}\frac{\partial}{\partial t}\text{\textbf{.}}
\label{ct}%
\end{equation}
$\hat{H}$ is to be operated as in%
\begin{equation}
\left(
\begin{array}
[c]{cc}%
\phi_{r} & \phi_{c}%
\end{array}
\right)  \left(  c_{t}\text{\textbf{J}}\frac{\partial}{\partial t}\right)
\left(
\begin{array}
[c]{c}%
\phi_{r}\\
\phi_{c}%
\end{array}
\right)  =\left(
\begin{array}
[c]{cc}%
\phi_{r} & \phi_{c}%
\end{array}
\right)  \hat{H}\left(
\begin{array}
[c]{c}%
\phi_{r}\\
\phi_{c}%
\end{array}
\right)  . \label{ct2}%
\end{equation}
The constant $c_{t}$ along with the explicit form of $\hat{H}$ will be
determined later.

\subsubsection{The equation of continuity for\ $\rho(q,t)=\bar{\psi}%
(q,t)^{T}\bar{\psi}(q,t)$}

The conservation of $\rho(q,t)$ or the incompressibility thereof naturally
leads to the equation of continuity%
\begin{equation}
\frac{\partial}{\partial t}\rho(q,t)=-\nabla\cdot\vec{j}(q,t)
\label{Econtinuity}%
\end{equation}
with the flux being defined%
\begin{equation}
\vec{j}=\vec{v}\rho,
\end{equation}
where $\vec{v}(q,t)$ indicates the local velocity at space-time. Since the
momentum at $(q,t)$ is given by Eq. (\ref{p-op2}), the relevant flux should be
defined as%
\begin{align}
\vec{j}  &  =\left(
\begin{array}
[c]{cc}%
\phi_{r}(q,t) & \phi_{c}(q,t)
\end{array}
\right)  \frac{\hat{p}}{m}\left(
\begin{array}
[c]{c}%
\phi_{r}(q,t)\\
\phi_{c}(q,t)
\end{array}
\right) \nonumber\\
&  =\frac{\hbar}{m}\left(  \phi_{r}\vec{\nabla}\phi_{c}-\phi_{c}\vec{\nabla
}\phi_{r}\right)  . \label{primeflux}%
\end{align}
The continuity equation (\ref{Econtinuity}), which is also referred to as the
law of flux conservation, is rewritten as
\begin{align}
&  \left(
\begin{array}
[c]{cc}%
\phi_{r} & \phi_{c}%
\end{array}
\right)  \frac{\partial}{\partial t}\left(
\begin{array}
[c]{c}%
\phi_{r}\\
\phi_{c}%
\end{array}
\right)  =-\frac{\hbar}{2m}\vec{\nabla}\cdot\left(  \phi_{r}\vec{\nabla}%
\phi_{c}-\phi_{c}\vec{\nabla}\phi_{r}\right) \nonumber\\
&  =-\frac{\hbar}{2m}\left(
\begin{array}
[c]{cc}%
\phi_{r} & \phi_{c}%
\end{array}
\right)  \vec{\nabla}^{2}\left(
\begin{array}
[c]{c}%
\phi_{c}\\
-\phi_{r}%
\end{array}
\right) \nonumber\\
&  =\left(
\begin{array}
[c]{cc}%
\phi_{r} & \phi_{c}%
\end{array}
\right)  \left(  -\text{\textbf{J}}\frac{\hat{p}^{2}}{2m\hbar}\right)  \left(
\begin{array}
[c]{c}%
\phi_{r}\\
\phi_{c}%
\end{array}
\right)  , \label{flux0}%
\end{align}
where we have used%
\begin{equation}
\hat{p}^{2}=-c_{p}^{2}\text{\textbf{J}}^{2}\nabla^{2}=c_{p}^{2}\nabla
^{2}\text{\textbf{I}}. \label{kinene}%
\end{equation}
\newline

\subsubsection{Meeting with the quantum phenomena}

The present derivation towards the Schr\"{o}dinger equation up to the next
treatment of the variational principle is universal in that only the three
general requirements are imposed on the Schr\"{o}dinger vector representation
of the density, Eq. (\ref{fact}). Therefore there is no physical ground on
which to determined the arbitrary constant $c_{p}$ within the present frame,
and it can be determined only by comparison with quantum experiments. The
historical experiments having opened the age of quantum physics had identified
the Planck constant $\hbar$ as a universal constant prior to the theoretical
developments,\cite{dirac1981principles,Schiff,Bohm-text,
Messiah,schwinger2001quantum,Feynman-Hibbs,ruetsche2011interpreting,omnes1992consistent}
and we already know $c_{p}^{2}$ in Eq. (\ref{kinene}) has to be equal to
$\hbar^{2}$ in order to meet the quantum phenomena. Just as a little more
precise example, we consider a \textquotedblleft pure
distribution\textquotedblright\ in classical dynamics\
\begin{equation}
\rho(q,t)=1.0\text{ in }q\in\left[  q_{1}(t),q_{2}(t)\right]  , \label{exrho}%
\end{equation}
with $q_{i}(t)=q_{i}(t_{0})+p_{0}(t-t_{0})/m$ $\ (i=1,2)$ and $m~$being a
mass. In this specific example, we may choose $\bar{\psi}(q,t),$ among others,
with no loss of generality as
\begin{equation}
\phi_{r}(q,t)=\cos(\frac{p_{0}}{K}q+\chi)\text{ \ \ and \ }\phi_{c}%
(q,t)=\sin(\frac{p_{0}}{K}q+\chi), \label{pln}%
\end{equation}
in $q\in\left[  q_{1}(t),q_{2}(t)\right]  $ and zero otherwise, since
$\phi_{r}^{2}(q,t)+$ $\phi_{c}^{2}(q,t)=\rho(q,t)=1.0$. $K$ is an arbitrary
constant having the dimension of action, $\chi$ being an arbitrary constant.
Then, Eq. (\ref{pln}) gives rise to%

\begin{equation}
\left(
\begin{array}
[c]{cc}%
\phi_{r}(q,t) & \phi_{c}(q,t)
\end{array}
\right)  \hat{p}\left(
\begin{array}
[c]{c}%
\phi_{r}(q,t)\\
\phi_{c}(q,t)
\end{array}
\right)  =-\frac{p_{0}}{K}c_{p}. \label{plane}%
\end{equation}
It is natural to regard that the value of the left hand side is equivalent to
$p_{0}$, and $-\frac{p_{0}}{K}c_{p}=p_{0}$ results in $c_{p}=-K.$ Further, the
quantum experiments like the Compton scattering\cite{Messiah} demands that
\begin{equation}
K=-c_{p}=\hbar
\end{equation}
and we have%
\begin{equation}
\hat{p}=-\hbar\text{\textbf{J}}\nabla. \label{momentum}%
\end{equation}

\subsection{The real-valued Schr\"{o}dinger equation to pick the most likely
state \label{sec:variational}}

\subsubsection{Case of free field}

We next consider the following variational condition to take the most likely
$\rho(q,t)$ out of those that satisfy Eq. (\ref{flux0}) as%

\begin{equation}
\left(
\begin{array}
[c]{cc}%
\phi_{r}+\delta\phi_{r} & \phi_{c}+\delta\phi_{c}%
\end{array}
\right)  \left(  \frac{\partial}{\partial t}+\text{\textbf{J}}\frac{\hat
{p}^{2}}{2m\hbar}\right)  \left(
\begin{array}
[c]{c}%
\phi_{r}+\delta\phi_{r}\\
\phi_{c}+\delta\phi_{c}%
\end{array}
\right)  =0. \label{vp0}%
\end{equation}
The left variation gives rise to
\begin{equation}
\left(
\begin{array}
[c]{cc}%
\delta\phi_{r} & \delta\phi_{c}%
\end{array}
\right)  \left(  \hbar\frac{\partial}{\partial t}+\text{\textbf{J}}\frac
{\hat{p}^{2}}{2m}\right)  \left(
\begin{array}
[c]{c}%
\phi_{r}\\
\phi_{c}%
\end{array}
\right)  =0, \label{vp1}%
\end{equation}
along with the right counterpart. Due to the arbitrariness of $\left(
\begin{array}
[c]{cc}%
\delta\phi_{r} & \delta\phi_{c}%
\end{array}
\right)  $, the dynamics of $\bar{\psi}(q,t)$ arises as%
\begin{equation}
\hbar\frac{\partial}{\partial t}\left(
\begin{array}
[c]{c}%
\phi_{r}\\
\phi_{c}%
\end{array}
\right)  =-\text{\textbf{J}}\frac{\hat{p}^{2}}{2m}\left(
\begin{array}
[c]{c}%
\phi_{r}\\
\phi_{c}%
\end{array}
\right)  .
\end{equation}
Or, multiplication of \textbf{J }on the both sides rewrite it to the prototype
of the familiar Schr\"{o}dinger equation
\begin{equation}
\hbar\text{\textbf{J}}\frac{\partial}{\partial t}\left(
\begin{array}
[c]{c}%
\phi_{r}\\
\phi_{c}%
\end{array}
\right)  =\frac{\hat{p}^{2}}{2m}\left(
\begin{array}
[c]{c}%
\phi_{r}\\
\phi_{c}%
\end{array}
\right)  \label{Sch0}%
\end{equation}
in the present representation.

\subsubsection{Case of the presence of the finite potential function}

Before proceeding, we note the notion of skew orthogonality, which is%
\begin{equation}
\left(
\begin{array}
[c]{cc}%
\phi_{r} & \phi_{c}%
\end{array}
\right)  \text{\textbf{J}}\left(
\begin{array}
[c]{c}%
\phi_{r}\\
\phi_{c}%
\end{array}
\right)  =0.
\end{equation}
Back in Eq. (\ref{flux0}), we find that the skew orthogonality keeps it
invariant as%
\begin{align}
&  \left(
\begin{array}
[c]{cc}%
\phi_{r} & \phi_{c}%
\end{array}
\right)  \left(  \hbar\frac{\partial}{\partial t}\right)  \left(
\begin{array}
[c]{c}%
\phi_{r}\\
\phi_{c}%
\end{array}
\right) \nonumber\\
&  =\left(
\begin{array}
[c]{cc}%
\phi_{r} & \phi_{c}%
\end{array}
\right)  \left(  -\text{\textbf{J}}\frac{\hat{p}^{2}}{2m}+\text{\textbf{J}%
}f(q,t)\right)  \left(
\begin{array}
[c]{c}%
\phi_{r}\\
\phi_{c}%
\end{array}
\right)  , \label{Sch1}%
\end{align}
where $f(q,t)$ is an arbitrary scalar function. A natural choice of $f(q,t)$
in our quantum dynamics is
\begin{equation}
f(q,t)=-V(q,t), \label{fchoice}%
\end{equation}
knowing that the potential energy function $V(q,t)$ reproduces the
Hamiltonian
\begin{equation}
\hat{H}=\frac{\hat{p}^{2}}{2m}+V(q,t). \label{Hamil}%
\end{equation}
Thus the variational condition is extended such that%
\begin{align}
&  \left(
\begin{array}
[c]{cc}%
\delta\phi_{r} & \delta\phi_{c}%
\end{array}
\right)  \left(  \hbar\frac{\partial}{\partial t}\right)  \left(
\begin{array}
[c]{c}%
\phi_{r}\\
\phi_{c}%
\end{array}
\right) \nonumber\\
&  =\left(
\begin{array}
[c]{cc}%
\delta\phi_{r} & \delta\phi_{c}%
\end{array}
\right)  \text{\textbf{J}}\left(  \frac{\hat{p}^{2}}{2m}+V(q,t)\right)
\left(
\begin{array}
[c]{c}%
\phi_{r}\\
\phi_{c}%
\end{array}
\right)  . \label{var}%
\end{align}
The variational principle after all demands the following dynamics
\begin{equation}
\hbar\frac{\partial}{\partial t}\left(
\begin{array}
[c]{c}%
\phi_{r}\\
\phi_{c}%
\end{array}
\right)  =-\text{\textbf{J}}\hat{H}\left(
\begin{array}
[c]{c}%
\phi_{r}\\
\phi_{c}%
\end{array}
\right)  , \label{Sch3}%
\end{equation}
or
\begin{align}
\hbar\text{\textbf{J}}\frac{\partial}{\partial t}\left(
\begin{array}
[c]{c}%
\phi_{r}\\
\phi_{c}%
\end{array}
\right)   &  =\hat{H}\left(
\begin{array}
[c]{c}%
\phi_{r}\\
\phi_{c}%
\end{array}
\right) \nonumber\\
&  =\left(  -\frac{\hbar^{2}}{2m}\nabla^{2}+V\right)  \left(
\begin{array}
[c]{c}%
\phi_{r}\\
\phi_{c}%
\end{array}
\right)  . \label{Sch30}%
\end{align}
We have now determined the energy operator $\hat{H}$ in Eq. (\ref{ct}) and
simultaneously determine $c_{t}=1$.

It should be emphasized again that since $\rho(q,t)$ is quadratic in
$\bar{\psi}(q,t)$, the stationary state point of the above variational
functional is located at the state where the relevant density appears to be
most likely in the functional space. Therefore $\bar{\psi}(q,t)$ satisfying
the real-valued Schr\"{o}dinger equation should optimizes the flux of the
density so as to orchestrate the possible internal states involved in
$\rho(q,t)$ in a coherent manner. \ 

The matrix operators of $\hat{H}$ and \ $\hat{p}$ are summarized as
\begin{equation}
\hat{H}=\text{\textbf{J}}\hbar\frac{\partial}{\partial t}\text{ \ \ \ \ and
\ }\hat{p}=-\text{\textbf{J}}\hbar\vec{\nabla} \label{H}%
\end{equation}
to be applied to the real-valued Schr\"{o}dinger vector.

\subsection{Some relevant properties of the real-valued Schr\"{o}dinger
equation}

\subsubsection{The Heisenberg dynamics}

One can move from the Schr\"{o}dinger dynamics to the operator algebra. For
the later purpose, we here touch only on the Heisenberg equation. For an
arbitrary time-independent operator $\hat{A}$ it follows that
\begin{align}
&  \frac{\partial}{\partial t}\left(
\begin{array}
[c]{cc}%
\phi_{r}(q,t) & \phi_{c}(q,t)
\end{array}
\right)  \hat{A}\left(
\begin{array}
[c]{c}%
\phi_{r}(q,t)\\
\phi_{c}(q,t)
\end{array}
\right) \nonumber\\
&  =-\frac{1}{\hbar}\left(
\begin{array}
[c]{cc}%
\phi_{r} & \phi_{c}%
\end{array}
\right)  \left(  \text{\textbf{J}}^{T}\hat{H}\hat{A}+\hat{A}\hat
{H}\text{\textbf{J}}\right)  \left(
\begin{array}
[c]{c}%
\phi_{r}\\
\phi_{c}%
\end{array}
\right) \nonumber\\
&  \equiv\frac{1}{\hbar}\left(
\begin{array}
[c]{cc}%
\phi_{r} & \phi_{c}%
\end{array}
\right)  \left[  \hat{H},\hat{A}\right]  \text{\textbf{J}}\left(
\begin{array}
[c]{c}%
\phi_{r}\\
\phi_{c}%
\end{array}
\right)  \label{Heisenberg}%
\end{align}
with $\left[  \hat{H},\hat{A}\right]  =\hat{H}\hat{A}-\hat{A}\hat{H}$, where
the Hamiltonian $\hat{H}$ of Eq. (\ref{Hamil}) is assumed to be Hermitian for
the square integrable (L$^{2}$) functions $\left(
\begin{array}
[c]{cc}%
\phi_{r} & \phi_{c}%
\end{array}
\right)  ^{T}.$ The extension to time dependent operators $\hat{A}(t)$ is
obvious. The conservation of energy is readily seen by putting $\hat{A}=$
$\hat{H}\,$.

\subsubsection{Time reversal equation}

We resume Eq. (\ref{Sch30}) with the form
\begin{equation}
\mathbf{P}_{3}\hbar\text{\textbf{J}}\frac{\partial}{\partial t}\left(
\begin{array}
[c]{c}%
\phi_{r}\\
\phi_{c}%
\end{array}
\right)  =\mathbf{P}_{3}\left(  -\frac{\hat{p}^{2}}{2m}+V(q,t)\right)  \left(
\begin{array}
[c]{c}%
\phi_{r}\\
\phi_{c}%
\end{array}
\right)  ,
\end{equation}
where
\begin{equation}
\mathbf{P}_{3}=\left(
\begin{array}
[c]{cc}%
1 & 0\\
0 & -1
\end{array}
\right)  .
\end{equation}
The relation
\begin{equation}
\mathbf{P}_{3}\text{\textbf{J}}=-\text{\textbf{J}}\mathbf{P}_{3},
\end{equation}
allows to further proceed to%
\begin{equation}
-\hbar\text{\textbf{J}}\frac{\partial}{\partial t}\mathbf{P}_{3}\left(
\begin{array}
[c]{c}%
\phi_{r}\\
\phi_{c}%
\end{array}
\right)  =\left(  -\frac{\hat{p}^{2}}{2m}+V(q,t)\right)  \mathbf{P}_{3}\left(
\begin{array}
[c]{c}%
\phi_{r}\\
\phi_{c}%
\end{array}
\right)  ,
\end{equation}
which reads%
\begin{equation}
-\hbar\text{\textbf{J}}\frac{\partial}{\partial t}\left(
\begin{array}
[c]{c}%
\phi_{r}(q,t)\\
-\phi_{c}(q,t)
\end{array}
\right)  =\left(  -\frac{\hat{p}^{2}}{2m}+V(q,t)\right)  \left(
\begin{array}
[c]{c}%
\phi_{r}(q,t)\\
-\phi_{c}(q,t)
\end{array}
\right)  , \label{conj}%
\end{equation}
and the replacement of $t\rightarrow-t$ gives the time-reversal form of the
Schr\"{o}dinger equation%

\begin{equation}
\hbar\text{\textbf{J}}\frac{\partial}{\partial t}\left(
\begin{array}
[c]{c}%
\phi_{r}(q,-t)\\
-\phi_{c}(q,-t)
\end{array}
\right)  =\left(  -\frac{\hat{p}^{2}}{2m}+V(q,-t)\right)  \left(
\begin{array}
[c]{c}%
\phi_{r}(q,-t)\\
-\phi_{c}(q,-t)
\end{array}
\right)  . \label{t-reverse}%
\end{equation}
Recall that the time reversal in a classical path $(q(t),p(t))$ is
$q_{rev}(-t)=q(t)$ and $p_{rev}(t)=-p(-t)$.

\subsubsection{Shift to the complex~number field: The canonical
Schr\"{o}dinger equation}

The basic properties of the standard symplectic matrix \textbf{J }is%

\begin{equation}
\text{\textbf{J}}^{2}=-\text{\textbf{I,}} \label{JJ}%
\end{equation}
and
\begin{equation}
\text{\textbf{J}}^{-1}=-\text{\textbf{J}}=\text{\textbf{J}}^{T}.
\end{equation}
Because of these properties, the symplectic form of two component vectors in
the real number field is essentially equivalent to the complex scalar field.
We may therefore set%

\begin{equation}
\text{\textbf{J}}\rightarrow i
\end{equation}
in the equations studied above thus far. For instance, the Hamiltonian and the
momentum in Eq. (\ref{H}) respectively should read%

\begin{equation}
\hat{H}=i\hbar\frac{\partial}{\partial t} \label{HH}%
\end{equation}
and
\begin{equation}
\hat{p}=-i\hbar\vec{\nabla}, \label{pp}%
\end{equation}
and should be operated on a scalar complex function%
\begin{equation}
\psi(q,t)=\phi_{r}(q,t)+i\phi_{c}(q,t), \label{Sf}%
\end{equation}
with the correspondence $\bar{\psi}(q,t)\leftrightarrow\psi(q,t)$. The
real-valued vector Schr\"{o}dinger equation (\ref{Sch3}) now appears to be the
canonical Schr\"{o}dinger equation%
\begin{equation}
i\hbar\frac{\partial}{\partial t}\psi(q,t)=\left(  -\frac{\hbar^{2}}{2m}%
\nabla^{2}+V\right)  \psi(q,t). \label{Sch4}%
\end{equation}
Conversely, the real and imaginary parts of Eq. (\ref{Sch4}) give back the
real-valued Schr\"{o}dinger equation Eq. (\ref{Sch30}). The expression of Eq.
(\ref{t-reverse}) naturally gives the time-reversal counterpart%

\begin{equation}
i\hbar\frac{\partial}{\partial t}\psi^{\ast}(q,-t)=\hat{H}\psi^{\ast}(q,-t).
\label{Sch5}%
\end{equation}

We now know how and why the Schr\"{o}dinger equation demands the imaginary
number in it. We note on the other hand that the complex-valued equation of
motion, Eq. (\ref{Sch4}) is far easier to handle both mathematically and
numerically. Also the treatment of the further complicated dynamics like the
relativistic one does not go through well without the complex algebra.
Nevertheless, the quantum dynamics in the present representation is insightful
and instructive to facilitate deeper understanding what the Schr\"{o}dinger
equation and functions are.

\subsubsection{Velocity and energy fields with vector rotation
\label{sec:velocityField}}

The Schr\"{o}dinger vector $\bar{\psi}(q,t)$ is normalizable such that%
\begin{equation}
\int dq\bar{\psi}(q,t)^{T}\wedge\left(  \text{\textbf{J}}\bar{\psi
}(q,t)\right)  =1, \label{psi-wedge}%
\end{equation}
which reminds of the Poincar\'{e}-Cartan theorem of integral invariance in
classical mechanics.\cite{Arnold} There are basically two ways of
normalization for physical quantities. One is the usual average (expectation
value), for instance,%
\begin{equation}
E^{\text{av}}=\left\langle \hat{H}\right\rangle =\frac{\int dq\left(
\begin{array}
[c]{cc}%
\phi_{r}(q,t) & \phi_{c}(q,t)
\end{array}
\right)  \hat{H}\left(
\begin{array}
[c]{c}%
\phi_{r}(q,t)\\
\phi_{c}(q,t)
\end{array}
\right)  }{\int\rho(q,t)dq}.
\end{equation}
The other is a local space-time distribution of the corresponding quantity as%
\begin{align}
E^{\text{local}}(q,t)  &  =\frac{1}{\rho}\left(
\begin{array}
[c]{cc}%
\phi_{r}(q,t) & \phi_{c}(q,t)
\end{array}
\right)  \hat{H}\left(
\begin{array}
[c]{c}%
\phi_{r}(q,t)\\
\phi_{c}(q,t)
\end{array}
\right) \nonumber\\
&  =\operatorname{Re}\frac{\hat{H}\psi}{\psi}. \label{e-density}%
\end{align}
Likewise the space-time distribution of the velocity is%

\begin{align}
\vec{v}^{\text{local}}(q,t)  &  =\frac{1}{\rho}\left(
\begin{array}
[c]{cc}%
\phi_{r}(q,t) & \phi_{c}(q,t)
\end{array}
\right)  \frac{\hat{p}}{m}\left(
\begin{array}
[c]{c}%
\phi_{r}(q,t)\\
\phi_{c}(q,t)
\end{array}
\right) \nonumber\\
&  =\frac{\hbar}{\rho m}\left(  \phi_{r}\vec{\nabla}\phi_{c}-\phi_{c}%
\vec{\nabla}\phi_{r}\right) \nonumber\\
&  =\frac{\hbar}{m}\operatorname{Im}\frac{\vec{\nabla}\psi}{\psi}.
\label{v-density}%
\end{align}
These quantities are homogeneous of degree zero in $\bar{\psi}$. The local
velocity $\vec{v}^{\text{local}}(q,t)$ of Eq. (\ref{v-density}) will appear in
the later study on the relationship between the Feynman-Kac formula, the
diffusion equation, and Ito stochastic differential equation later in Sec.
\ref{sec:Stochastic}.

We next consider the physical meaning of the local velocity of Eq.
(\ref{v-density}) and local energy of Eq. (\ref{e-density}). Let us write the
Schr\"{o}dinger vector in a polar coordinate with the length and rotation
angle such that
\begin{equation}
\bar{\psi}(q,t)^{T}=\rho(q,t)^{1/2}(\cos\theta(q,t)\text{ \ }\sin\theta(q,t)),
\label{qdist}%
\end{equation}
which is equivalent to the complex-valued Schr\"{o}dinger function represented
as
\begin{equation}
\psi(q,t)=\rho(q,t)^{1/2}\exp\left(  i\theta(q,t)\right)
\end{equation}
as in the Bohm
representation.\cite{bohm1952suggested,Bohm-text,Wyatttext,sanz2013trajectory}
(Note that the Bohm representation is not about the derivation of the
Schr\"{o}dinger equation but its alternative expression.) Then it turns that%

\begin{equation}
\vec{v}^{\text{local}}(q,t)=\frac{\hbar}{m}\vec{\nabla}\theta(q,t)
\label{vangle}%
\end{equation}
and%
\begin{equation}
E^{\text{local}}(q,t)=\hbar\partial_{t}\theta(q,t). \label{Eangle}%
\end{equation}
Therefore, the local velocity is essentially equivalent to the rate of the
vector rotation in $q$ space at a given time, while the local energy is
proportional to the speed of angular rotation at a given $q$. Thus a particle
at a point $q$ is to be sent forward to a next point with the velocity of the
local rotation.\ Indeed, $\vec{v}^{\text{local}}(q,t)$ will appear in the next
section as a drift velocity to drive a quantum stochastic dynamics. Therefore
it is rational to imagine that an implicit \textquotedblleft vector
rotation\textquotedblright\ is equipped as an intrinsic machinery in the
Schr\"{o}dinger dynamics. It may be referred to as \textquotedblleft internal
rotation\textquotedblright\ of the Schr\"{o}dinger vector. Equation
(\ref{Eangle}) shows that the stationary states are materialized when the
conditions%
\begin{equation}
E^{\text{local}}(q,t)=\hbar\partial_{t}\theta(q,t)=E\text{ \ \ \ (constant)}%
\end{equation}
with%
\begin{equation}
\hat{H}\psi=E\psi.
\end{equation}

Equations (\ref{vangle}) and (\ref{Eangle}), suggest that $\hbar\theta(q,t)$
looks similar to the classical action function.\cite{Arnold} This fact partly
underlies the \textquotedblleft derivation\textquotedblright\ of the
Schr\"{o}dinger equation by himself. We will study the role of $\theta(q,t)$
in a great detail in the context of the dynamics of quantum stochastic paths
and the associated quantum canonical equations of motion in Secs.
\ref{sec:Stochastic} and \ref{sec:correlation}.

\subsection{$\rho(q,t)$ and $\bar{\psi}(q,t)$}

The density $\rho(q,t)$ in quantum mechanics is generally a mixture of
innumerable \textquotedblleft possible physical phenomena\textquotedblright,
which cannot necessarily be unfolded into independent ones in principle.
Likewise, $\psi(q,t)$ represents a \textquotedblleft
coherent\textquotedblright\ distribution amplitude for an ensemble of
theoretically possible events to happen in quantum level and is natural to be
referred to as quantum distribution amplitude function. It is therefore
inappropriate to regard that a single Schr\"{o}dinger function describes a
singly isolated phenomenon or event. It is also wrong to regard a
Schr\"{o}dinger function as a dynamical function materializing a physical
substance. Instead, the variational principle leading to the Schr\"{o}dinger
equation allows to regard $\bar{\psi}(q,t)$ as a \textquotedblleft back-ground
mathematical machinery\textquotedblright\ to materialize the most likely
$\rho(q,t)$ among those satisfying the space-time translational invariance and
the flux conservation under a given initial condition. Thus the
Schr\"{o}dinger equation is not a axiom. It bears the mechanical role and function.

As for the wave-particle duality, it is now widely denied to regard
$\psi(q,t)$ as a representation of a wave of any physical
substance.\cite{auletta2001foundations,home2007einstein,whitaker2012new,freire2022oxford}
The wave-like nature of $\psi(q,t)$ emerges from the linearity of the
Schr\"{o}dinger equation. The linearity in turn naturally brings about the
superposition principle. However, it should be noted that the linearity does
not always lead to the properties deduced from the Huygens principle, such as
diffraction, bifurcation, specific interference, and so on. As for the
Huygens-like principle in the Schr\"{o}dinger wavepacket dynamics, we refer to
ref. \onlinecite{Goussev-Huygens,takatsuka2023schrodinger}.

We confirm that the dynamics of the Schr\"{o}dinger vector $\bar{\psi}(q,t)$
does not have the internal mechanism of instantaneous collapse. (See, however,
ref. \onlinecite{bassi2013models} for a review on this matter in different
perspectives.) It is well known that the semiclassical wavepacket can diverge
to a delta-function at caustics and turning points. However, the mechanism of
quantum smoothing of those divergences are well analyzed\cite{Paper-II} and
the seeming divergence has nothing to do with the instantaneous collapse
claimed by the so-called Copenhagen
interpretation.\cite{auletta2001foundations,home2007einstein,whitaker2012new,freire2022oxford}
As will be discussed in the next section, the quantum path dynamics denies the
necessity of the notion of collapse in the reality of dynamics.

\section{Quantum path integrals in the real-valued configuration
space\label{sec:Feynman-Kac}}

Now that the real-valued Schr\"{o}dinger equation is available at hand, we
proceed to find the real-valued quantum path integrals as the Green function
of it. Then, the comparison with the Feynman-Kac formula in statistical
mechanics will suggest the existence of the dynamics of stochastic paths
behind quantum mechanics.

\subsection{The Feynman path integrals}

We first revisit the Feynman path
integrals,\cite{Feynman,Feynman-Hibbs,Schulman,kleinert2006path} which is%

\begin{align}
&  K(q,t)=\left\langle q\left\vert \exp\left(  \frac{1}{i\hbar}\hat
{H}t\right)  \right\vert 0\right\rangle \nonumber\\
&  =\lim_{N\rightarrow\infty}\int d^{3}q_{1}\cdots\int d^{3}q_{N}\left(
\frac{m}{2\pi i\hbar\Delta t}\right)  ^{3(N+1)/2}\nonumber\\
&  \times\exp\left[  \frac{i}{\hbar}\sum_{k=0}^{N}\left(  \frac{m}{2}%
\frac{\left(  q_{k+1}-q_{k}\right)  ^{2}}{\Delta t}-V(q_{k})\Delta t\right)
\right]  \label{Fkernel}%
\end{align}
with $\Delta t=t/(N+1)$. As is well known, the kernel $K(q,t)$ demands the
democratic summation of continuous polylines (broken lines), each connecting
two neighboring positions $q_{k}$ and $q_{k+1}$ of an infinitessimal distance.
There is no need to stress the profound value of this path representation in
quantum sciences. This kernel seems somewhat similar to the Wiener
path-integrals for the Brownian motion (shown later in order in Eq.
(\ref{WienerG})), but it is not really the case, in that the integral measure
is not well prepared in Eq. (\ref{Fkernel}),\cite{Klauder1,Klauder2} and the
convergence with respect to the number of sampling paths is not mathematically
secured. The paths are pure mathematical quantities and are not subject to any
dynamics. Therefore each path may be regarded as a \textquotedblleft basis
function\textquotedblright\ to expand the kernel.\cite{davison1954feynmann}
The coherent interference among those paths is the very core of the theory,
and for example, the stationary-phase approximation extracts classical
trajectories leading to a semiclassics
mechanics.\cite{Messiah,Schulman,kleinert2006path}\color{black}

\subsection{The real-valued path integrals}

We next briefly outline the path integrals in statistical mechanics. The time
propagation of the diffusion equation of a diffusion constant $D$
\begin{equation}
\frac{\partial}{\partial t}\Phi(q,t)=\left(  D\nabla^{2}-\lambda V(q)\right)
\Phi(q,t) \label{Bdiff}%
\end{equation}
is well known to have the following coordinate representation as
\begin{align}
&  \Phi(x,t+\Delta t)=\frac{1}{\left(  4\pi D\Delta t\right)  ^{1/2}%
}\nonumber\\
&  \times\int_{-\infty}^{\infty}dy\exp\left[  -\Delta t\left(  \frac{1}%
{4D}\left(  \frac{x-y}{\Delta t}\right)  ^{2}+\lambda V(y,t)\right)  \right]
\Phi(y,t),
\end{align}
and accordingly the path integral representation for a finite time
propagation\cite{kac1949distributions,del2004feynman,Ezawa} gives the Green
function of Eq. (\ref{Bdiff}), which is
\begin{align}
&  G(q,t)\nonumber\\
&  =\int_{\Omega\left(  q,t:0.0)\right)  }\exp\left[  -\lambda\int_{0}%
^{t}V(s,X_{s}(\omega))ds\right]  dP_{W\left(  q,t:0.0)\right)  }(\omega),
\label{FC}%
\end{align}
where $\Omega\left[  q,t:0,0\right]  $ is a set of sampling paths reaching
from $\left(  0,0\right)  $ and arriving at $\left(  q,t\right)  $, and
$\omega$ specifies as a member of $\Omega$, and $dP_{W\left(  q,t:0.0)\right)
}(\omega)$ is the Wiener measure of the Brownian motion, or more explicitly%

\begin{align}
G(q,t)  &  =\lim_{N\rightarrow\infty}\frac{1}{\left(  4\pi D\Delta t\right)
^{N/2}}\prod_{k=1}^{\infty}\int_{-\infty}^{\infty}dq_{k}\nonumber\\
&  \times\exp\left[  -\sum_{k=0}^{N-1}\Delta t\left(  \frac{1}{4D}%
\frac{\left(  \Delta q_{k}\right)  ^{2}}{\Delta t}+\lambda V(q_{k}%
,t_{k})\right)  \right]  \label{WienerG}%
\end{align}
with $t_{k}=k\Delta t,$ $q_{k}=q(t_{k})$, $q_{0}=0$, \ $q_{n}=q,$ $\Delta
q_{k}=q_{k+1}-q_{k}$, and with $G(q,0)=\delta(q)$. The mathematical similarity
between $G(q,t)$ in Eq. (\ref{WienerG}) and the Feynman kernel of Eq.
(\ref{Fkernel}) is obvious. Indeed, it is told that Kac was inspired by
Feynman's path integrals to construct his formula.\cite{kac1949distributions}
However, the Feynman-Kac path integral is mathematically rigorous, but not
necessarily so is the Feynman path integration due to the lack of a
well-defined integral measure. This is because of the presence of the
imaginary number in each amplitude factor.

Our target is to find a path-integral representation of the real-valued
Schr\"{o}dinger equation Eq. (\ref{Sch30}), which is rewritten as%
\begin{equation}
\frac{\partial}{\partial t}\left(
\begin{array}
[c]{c}%
\phi_{r}(q,t)\\
\phi_{c}(q,t)
\end{array}
\right)  =\left(  \frac{\hbar}{2m}\nabla^{2}-\frac{V(q)}{\hbar}\right)
\text{\textbf{J}}\left(
\begin{array}
[c]{c}%
\phi_{r}(q,t)\\
\phi_{c}(q,t)
\end{array}
\right)  , \label{Sch45}%
\end{equation}
or shortly%

\begin{equation}
\frac{\partial}{\partial t}\bar{\psi}(q,t)=\left(  \frac{\hbar}{2m}\nabla
^{2}-\frac{1}{\hbar}V(q)\right)  \text{\textbf{J}}\bar{\psi}(q,t)\mathbf{.}%
\end{equation}
It is immediately noticed that the quantum dynamics Eq. (\ref{Sch45}) may be
regarded as a coupled diffusion dynamics with $D=\hbar/2m$ and $\lambda
=1/\hbar$ in the statistical counterpart Eq. (\ref{Bdiff}), and the coupling
constant is \textbf{J.} Equation (\ref{Sch45}) is formally integrated as%
\begin{equation}
\bar{\psi}(t+\Delta t)=\exp\left[  \left(  D\nabla^{2}-\lambda V(q)\right)
\text{\textbf{J}}\Delta t\right]  \bar{\psi}%
\end{equation}
for a short time $\Delta t.$ We take the first order expansion of this
exponential operator such that
\begin{align}
&  \bar{\psi}(t+\Delta t)\simeq\left(  \text{\textbf{I}}+\text{\textbf{J}%
}\left(  D\nabla^{2}-\lambda V(q)\right)  \Delta t\right)  \bar{\psi
}(t)\nonumber\\
&  \simeq\left(  \text{\textbf{I}}+\text{\textbf{J}}(\exp[\left(  D\nabla
^{2}-\lambda V(q)\right)  \Delta t\right)  ]-1))\bar{\psi}(t)\nonumber\\
&  \simeq\left(  \text{\textbf{I}}-\text{\textbf{J}}\right)  \bar{\psi
}(t)+\text{\textbf{J}}\exp\left[  \left(  D\nabla^{2}-\lambda V(q)\right)
\Delta t\right]  \bar{\psi}(t)\nonumber\\
&  =\left(  \text{\textbf{I}}-\text{\textbf{J}}\right)  \bar{\psi
}(t)+\text{\textbf{J}}\frac{1}{\left(  4\pi D\Delta t\right)  ^{1/2}%
}\nonumber\\
&  \times\int_{-\infty}^{\infty}dy\exp\left[  -\Delta t\left(  \frac{1}%
{4D}\left(  \frac{x-y}{\Delta t}\right)  ^{2}+\lambda V(y,t)\right)  \right]
\bar{\psi}(y,t), \label{P1}%
\end{align}
where \textbf{I }denotes the 2$\times$2unit matrix, and with $\Delta
t\rightarrow0$. Further, we may proceed with the Gaussian representation of
the Dirac delta function
\begin{equation}
\delta(x-y)=\frac{1}{\left(  4\pi D\Delta t\right)  ^{1/2}}\int_{-\infty
}^{\infty}dy\exp\left[  -\Delta t\left(  \frac{1}{4D}\left(  \frac{x-y}{\Delta
t}\right)  ^{2}\right)  \right]
\end{equation}
with $\Delta t\rightarrow+0$ to rewrite $\bar{\psi}(x,t)$ as%

\begin{align}
&  \bar{\psi}(x,t)=\int_{-\infty}^{\infty}dy\delta(x-y)\bar{\psi
}(y,t)\nonumber\\
&  =\frac{1}{\left(  4\pi D\Delta t\right)  ^{1/2}}\int_{-\infty}^{\infty
}dy\exp\left[  -\Delta t\left(  \frac{1}{4D}\left(  \frac{x-y}{\Delta
t}\right)  ^{2}\right)  \right]  \bar{\psi}(y,t). \label{P2}%
\end{align}
Putting Eq. (\ref{P2}) back into Eq. (\ref{P1}), we have%
\begin{align}
&  \bar{\psi}(x,t+\Delta t)\nonumber\\
&  =\frac{1}{\left(  4\pi D\Delta t\right)  ^{1/2}}\int_{-\infty}^{\infty
}dy\exp\left[  -\Delta t\left(  \frac{1}{4D}\left(  \frac{x-y}{\Delta
t}\right)  ^{2}\right)  \right] \nonumber\\
&  \times\left[  \left(  \text{\textbf{I}}-\text{\textbf{J}}\right)
+\text{\textbf{J}}\exp\left(  -\Delta t\lambda V(y,t)\right)  \right]
\mathbf{\Phi}(y,t)
\end{align}
\newline Once again, the exponential function is expanded to the first order
of $\Delta t$ and bring it back into another exponential form as%
\begin{align}
&  \left(  \text{\textbf{I}}-\text{\textbf{J}}\right)  +\text{\textbf{J}}%
\exp\left(  -\Delta t\lambda V(y,t)\right) \nonumber\\
&  \simeq\left(  \text{\textbf{I}}-\text{\textbf{J}}\right)  +\text{\textbf{J}%
}\left[  1-\Delta t\lambda V(y,t)\right] \nonumber\\
&  \simeq\text{\textbf{I}}-\left(  \Delta t\lambda V(y,t)\right)
\text{\textbf{J}}\simeq\exp\left[  -\Delta t\lambda V(y,t)\text{\textbf{J}%
}\right]  .
\end{align}
After all it results that
\begin{align}
&  \bar{\psi}(x,t+\Delta t)\nonumber\\
&  \simeq\frac{1}{\left(  4\pi D\Delta t\right)  ^{1/2}}\int_{-\infty}%
^{\infty}dy\exp\left[  -\Delta t\left(  \frac{1}{4D}\left(  \frac{x-y}{\Delta
t}\right)  ^{2}\right)  \right] \nonumber\\
&  \times\exp\left[  -\Delta t\lambda V(y,t)\text{\textbf{J}}\right]
\bar{\psi}(y,t)
\end{align}
to the first order of the very short $\Delta t.$ As usual, we repeat this
short time propagation to a finite time expression as%

\begin{align}
&  \mathbf{G}(q,t:0,0)=\lim_{N\rightarrow\infty}\int_{-\infty}^{\infty}\Delta
q_{1}\cdots\Delta q_{N-1}\left(  \frac{m}{2\pi\hbar\Delta t}\right)
^{N/2}\nonumber\\
&  \times\exp\left[  -\sum_{k=0}^{N-1}\frac{\Delta t}{\hbar}\left(  \frac
{m}{2}\left(  \frac{\Delta q_{k}}{\Delta t}\right)  ^{2}+V(q_{k}%
,t_{k})\text{\textbf{J}}\right)  \right]  , \label{matrixGreen}%
\end{align}
with $\Delta q_{k}=q_{k+1}-q_{k}$, $q_{N}=q$, $q_{0}=0$, returning to
$D\rightarrow\hbar/2m$, \ \ $\lambda\rightarrow1/\hbar$. We thus define the
Green function for Eq. (\ref{Sch45}) and%
\begin{align}
\mathbf{G}(q,t  &  :0,0)=\int_{\Omega\left[  q,t:0,0\right]  }\exp\left[
-\frac{1}{\hbar}\int_{0}^{t}V\left(  s,X\left(  s,\omega\right)  \right)
ds)\text{\textbf{J}}\right] \nonumber\\
&  \times dP_{W\left[  q,t:0,0\right]  }(\omega) \label{PI-real}%
\end{align}
with the Wiener measure%
\begin{align}
&  dP_{W[q(s+\Delta s,\omega),s+\Delta s:q(s,\omega),s]}(\omega)=dq\left(
\frac{m}{2\pi\hbar\Delta s}\right)  ^{N/2}\nonumber\\
&  \exp\left[  -\Delta s\left(  \frac{m}{2\hbar}\left(  \frac{q(s+\Delta
s,\omega)-q(s,\omega)}{\Delta s}\right)  ^{2}\right)  \right]  . \label{WM}%
\end{align}
Note that the standard symplectic matrix \textbf{J} is associated only with
the potential function $V(q_{k},t_{k})$ but is not involved in the Wiener
measure. After all it holds%
\begin{equation}
\left(
\begin{array}
[c]{c}%
\phi_{r}(q,t)\\
\phi_{c}(q,t)
\end{array}
\right)  =\int dq\mathbf{G}(q,t:q_{0},0)\left(
\begin{array}
[c]{c}%
\phi_{r}(q_{0},0)\\
\phi_{c}(q_{0},0)
\end{array}
\right)  . \label{Gprop}%
\end{equation}

Equations (\ref{matrixGreen}) or formally equivalent (\ref{PI-real}) is an
extension of the Feynman-Kac formula to the system of coupled diffusion
equation, Eq. (\ref{Sch45}), and a real-valued realization of the Feynman path
integrals. In the Feynman path integration each path bears a finite amplitude
arising from $\left(  \frac{m}{2\pi i\hbar\Delta t}\right)  ^{3(N+1)/2}$ as in
Eq. (\ref{Fkernel}) and the most of unphysical paths are to be cancelled in
the summation over the highly oscillatory phases (according to the
Riemann-Lebesgue lemma). Hence even paths that break the relativity limit are
\textquotedblleft mathematically\textquotedblright\ allowed. It is thus hard
to make a clear-cut statement that the path integration in complex number
space practically converge as the sampled paths are added one by one into the
summation, unless additional conditions or mathematical tricks are
imposed.\cite{Klauder1,Klauder2} In\ the Wiener measure of Eq. (\ref{WM}), on
the other hand, the contribution from those paths are nullified automatically,
and the Green function of Eq. (\ref{PI-real}) secures the convergence in the
path summation.

The most significant aspect of Eqs. (\ref{matrixGreen}) and (\ref{PI-real}) to
the present work is that \textquotedblleft scalar and stochastic
paths\textquotedblright\ should serve as a trail on which to carry the
Schr\"{o}dinger vectors. This is not obvious without proof. Moreover, the
present extension of the Feynman-Kac formula suggests that there should exist
stochastic paths behind the real-valued Schr\"{o}dinger equation. In the next
section, we find the actual quantum paths in an Ito stochastic differential equation.

\section{Quantum stochastic path dynamics \label{sec:Stochastic}}

We study the quantum stochastic path dynamics in this section as the second
pillar of the Schr\"{o}dinger dynamics.

\subsection{Dynamical path concepts}

We first briefly review three quantum-path theories, which are relevant to the
present work.

\subsubsection{Nelson theory with the stochastic Newtonian equation}

Let us start from the theory of Nelson.\cite{Nelson,nelson2012review} The
outline is as follows. He first defines the forward and backward derivatives
of a position $q(t)$ with a \textquotedblleft conditional average
(expectation)\textquotedblright, denoted by $E_{t}[\cdot]$,\ such that
\begin{equation}
DX_{t}=\lim_{\Delta t\rightarrow+0}E_{t}\left[  \frac{q(t+\Delta
t)-q(t)}{\Delta t}\right]  \label{D1}%
\end{equation}
and%
\begin{equation}
D_{\ast}X_{t}=\lim_{\Delta t\rightarrow+0}E_{t}\left[  \frac{q(t)-q(t-\Delta
t)}{\Delta t}\right]  , \label{D2}%
\end{equation}
which are naturally applied to define the mean forward velocity and backward
velocity
\begin{equation}
DX_{t}=b(X_{t},t)\text{ \ and }D_{\ast}X_{t}=b_{\ast}(X_{t},t), \label{drift1}%
\end{equation}
where $X_{t}$ indicates the stochastic variable corresponding to the position
$q$. Then $b(X_{t},t)$ and $b_{\ast}(X(t),t)$ are naturally regarded as the
velocity drift terms in the stochastic differential
equations\cite{van1976stochastic,risken1996fokker,Gardiner,oksendal2013stochastic}
\begin{equation}
dX_{t}=b(X_{t},t)dt+dW \label{st1}%
\end{equation}
and%
\begin{equation}
dX_{t\ast}=b_{\ast}(X_{t},t)dt+dW_{\ast} \label{st2}%
\end{equation}
respectively, with $W$ and $W_{\ast}$ being the Wiener process. Remarkable is
his definition of the (stochastic) Newtonian equation%
\begin{equation}
\frac{1}{2}(DD_{\ast}+D_{\ast}D)X_{t}=-\frac{1}{m}\frac{\partial V}{\partial
q}, \label{Nelson}%
\end{equation}
which has thus a time-reversal symmetry. After some manipulation, including
nonlinear equations on the way, he successfully arrived at the time-dependent
Schr\"{o}dinger equation (not reproduced here). Nelson claims that the
Schr\"{o}dinger equation could be derived only with \textquotedblleft
classical mechanics\textquotedblright\ of Eq. (\ref{Nelson}) and that the
probabilistic nature of the Schr\"{o}dinger dynamics naturally emerges through
the definitions of Eqs. (\ref{D1}) and (\ref{D2}). The title of his
paper\cite{Nelson} \textquotedblleft Derivation of the Schr\"{o}dinger
equation from Newton mechanics\textquotedblright\ suggests that the Newton
mechanics sets a foundation of quantum mechanics, or the both can be derived
on a single physical basis. However, we remind that the Schr\"{o}dinger
equation has been derived in such a straightforward and universal manner
without the help of other mechanics in Sec. \ref{sec:RealValue}.

Yasue sets his stochastic variational theory for the stochastic control to
derive the Schr\"{o}dinger equation.\cite{yasue1979stochastic,Yasue} He starts
from the following stationary condition of the energy%
\begin{equation}
\frac{d}{dt}\int\left[  \frac{1}{2}(\frac{m}{2}b^{2}+\frac{m}{2}b_{\ast}%
^{2})+V(q)\right]  \rho(q,t)=0.
\end{equation}
He also was successful in \textquotedblleft quantization\textquotedblright%
\ without use of the stochastic Newtonian equation Eq. (\ref{Nelson}).

\subsubsection{Nagasawa theory based on the Kolmogorov and Ito theorems}

Nagasawa's approach is much more fundamental than
Nelson's\cite{nagasawa1989transformations,Nagasawa2} in that he rests on the
rigorous theories in statistics by Kolmogorov and stochastic differential
equation by Ito. A brief outline is as follows. A Markov process for the
following parabolic partial differential equation (or a backward Fokker-Planck equation)%

\begin{equation}
\frac{\partial u}{\partial t}+\frac{1}{2}\sigma^{2}\nabla^{2}u+b(q,t)\cdot
\nabla u=0 \label{Kol}%
\end{equation}
is associated with the Ito stochastic process
\begin{equation}
X_{t}=X_{0}+\sigma B_{t}+\int_{0}^{t}b(X_{s},s)ds, \label{Kol2}%
\end{equation}
with $B_{t}$ being the Brownian process. Since the Schr\"{o}dinger equation is
complex, he figured out a set of real-valued equations satisfying Eq.
(\ref{Kol})
\begin{equation}
\frac{\partial\psi_{N}}{\partial t}+\frac{1}{2}\sigma^{2}\nabla^{2}\psi
_{N}+V(q,t)\psi_{N}=0 \label{Nagasawa1}%
\end{equation}
and
\begin{equation}
-\frac{\partial\tilde{\psi}_{N}}{\partial t}+\frac{1}{2}\sigma^{2}\nabla
^{2}\tilde{\psi}_{N}+V(q,t)\tilde{\psi}_{N}=0. \label{Nagasawa2}%
\end{equation}
$\psi_{N}$ and $\tilde{\psi}_{N}$ are a pair of real-valued solutions in the
functional forms
\begin{equation}
\psi_{N}(q,t)=\exp(R(q,t)+S(q,t))\text{ \ and \ }\tilde{\psi}_{N}=\exp(R-S),
\label{evolutionF}%
\end{equation}
with the real-valued functions $R(q,t)$ and $S(q,t)$ being the components of
the Schr\"{o}dinger function in the form
\begin{equation}
\psi=\exp\left(  R+iS\right)  , \label{Naga}%
\end{equation}
(note the position of $R$ in Eqs. (\ref{evolutionF}) and (\ref{Naga})).
Nagasawa referred to $\psi_{N}(x,t)$ and $\tilde{\psi}_{N}(x,t)$,
respectively, as the evolution function and the backward evolution function.
Note neither of them is the direct solution of the Schr\"{o}dinger equation.
Then, he showed that Eq. (\ref{Kol2}) gives the Ito stochastic process
\begin{equation}
dX_{t}=\frac{\hbar}{m}\nabla(R+S)dt+\sqrt{\frac{\hbar}{m}}dW, \label{Nagasawa}%
\end{equation}
where the velocity drift term is taken from the dynamics of Eq.
(\ref{Nagasawa1}). He has thus established the (indirect) relationship between
a stochastic theory and quantum dynamics. His equations Eqs. (\ref{Nagasawa1})
and (\ref{Nagasawa2}) are not the Schr\"{o}dinger equation, and thereby he
claims that his theory gives insights beyond the Schr\"{o}dinger
framework.\cite{nagasawa1989transformations,Nagasawa2} The Nagasawa theory is
truly monumental in that it clearly shows the stochasticity is mathematically
intrinsic and essential in quantum dynamics.

\subsection{Stochastic paths consistent with the Schr\"{o}dinger equation}

We now consider a quantum stochastic path dynamics in our own
way,\cite{takatsuka2025analysis} which is based on the relationship between
the Feynman-Kac formula for statistical physics, the corresponding diffusion
equation, and the Ito stochastic differential
equation.\cite{kac1949distributions,del2004feynman} The aim is to find the
quantum paths in the Schr\"{o}dinger dynamics.

\subsubsection{Triangle relation in stochastic dynamics: The diffusion
equation, the Feynman-Kac formula, and the stochastic differential equation}

To figure out the quantum stochastic paths compatible with the real-valued
Schr\"{o}dinger equation, we recall the triangle relationship between the
diffusion equation, the Feynman-Kac formula, and the stochastic differential
equation: (1) Eq. (\ref{Bdiff}) or
\begin{equation}
\frac{\partial\psi_{f}(q,t)}{\partial t}=\left(  D\nabla^{2}-\lambda
V(q,t)\right)  \psi_{f}(q,t) \label{Diff0}%
\end{equation}
represents the typical diffusion equation under a potential function. (2) The
Feynman-Kac formula, Eq. (\ref{FC}) or (\ref{WienerG}) serves as the Green
function of the diffusion differential equation. Both of these two are
concerned about the distribution of the stochastic dynamics on $V(q,t)$. (3)
Then, we need the path dynamics of the individual \textquotedblleft
Brownian\textquotedblright\ particle, which is represented in a stochastic
path $X_{t}=X\left(  t,\omega\right)  $ of a statistical sample $\omega$ in
the Ito differential equation%
\begin{equation}
dX_{t}=\alpha(X_{t},t)dt+dW\left(  t,\omega\right)  , \label{St1}%
\end{equation}
where $W\left(  t,\omega\right)  $ denotes the Wiener process. It has been
well established\cite{kac1949distributions,del2004feynman,Ezawa} that for the
stochastic differential equation Eq. (\ref{St1}) to be consistent with the
differential equation Eq. (\ref{Diff0}) and the Feynman-Kac formula Eq.
(\ref{WienerG}), the velocity drift term $\alpha(X_{t},t)$ must satisfy
\begin{equation}
\alpha(X_{t},t)=2D\frac{\nabla\psi_{f}}{\psi_{f}}. \label{velv1}%
\end{equation}
(The derivation of Eq. (\ref{velv1}) is not simple and is not reproduced
here.) We rewrite Eq. (\ref{St1}) formally as
\begin{equation}
dX_{t}=2D\frac{\nabla\psi_{f}}{\psi_{f}}dt+dW\left(  t,\omega\right)  ,
\label{FCdX}%
\end{equation}
which is the starting point towards the \textquotedblleft
quantum\textquotedblright\ stochastic path dynamics.

Here we remind of the quantum local velocity studied above in Eq.
(\ref{v-density})
\begin{equation}
v(q,t)=\frac{\hbar}{m}\operatorname{Im}\frac{\nabla\psi}{\psi} \label{velv2}%
\end{equation}
for the Schr\"{o}dinger function $\psi,$ which looks similar to Eq.
(\ref{velv1}). The comparison suggests a correspondence such that
\begin{equation}
D\leftrightarrow\hbar/2m. \label{diffconst}%
\end{equation}

\subsubsection{Quantum stochastic paths}

We now implant the information of the Schr\"{o}dinger equation into the
velocity drift term of the stochastic process, $\alpha(X_{t},t)$ of Eq.
(\ref{FCdX}). The real-valued Schr\"{o}dinger equation Eq. (\ref{Sch30}) in
the form of Eq. (\ref{Sch45}) seems to be fine to apply Eq. (\ref{Bdiff}).
However, Eq. (\ref{Sch45}) is actually composed of a pair of coupled
equations, and therefore we detour to formally uncouple them via the
complex-valued Schr\"{o}dinger equations as
\begin{equation}
i\frac{\partial}{\partial t}(\phi_{r}+i\phi_{c})=\left(  \frac{\hbar}%
{2m}\nabla^{2}-\frac{V}{\hbar}\right)  (\phi_{r}+i\phi_{c}) \label{Sch50}%
\end{equation}
and its complex conjugate%
\begin{equation}
-i\frac{\partial}{\partial t}(\phi_{r}-i\phi_{c})=\left(  \frac{\hbar}%
{2m}\nabla^{2}-\frac{V}{\hbar}\right)  (\phi_{r}-i\phi_{c}). \label{Sch51}%
\end{equation}
Equation (\ref{Sch50}) is further transformed to mimic Eq. (\ref{Diff0}) by
the rotation of time coordinate to%

\begin{equation}
\frac{\partial}{\partial s^{+}}\psi^{+}(q,s^{+})=\left(  \frac{\hbar}%
{2m}\nabla^{2}-\frac{V}{\hbar}\right)  \psi^{+}(q,s^{+}) \label{Sch40}%
\end{equation}
with%

\begin{equation}
s^{+}=-it \label{splus}%
\end{equation}
and $\psi^{+}=\phi_{r}+i\phi_{c}$. Another one comes from Eq. (\ref{Sch51})%

\begin{equation}
\frac{\partial}{\partial s^{-}}\psi^{-}(q,s^{-})=\left(  \frac{\hbar}%
{2m}\nabla^{2}-\frac{V}{\hbar}\right)  \psi^{-}(q,s^{-}) \label{Sch41}%
\end{equation}
by time rotation to the opposite direction $s^{+}$%

\begin{equation}
s^{-}=it \label{sminus}%
\end{equation}
and $\psi^{-}=\phi_{r}-i\phi_{c}$. The time rotations as in Eqs. (\ref{splus})
and (\ref{sminus}) are to be made at each time $t$. After the time rotations,
$\psi^{+}(q,s^{+})$ and $\psi^{-}(q,s^{-})$ are yet complex, resulting in the
complex valued velocity drift terms%
\begin{equation}
\alpha^{+}(X_{t},s^{+})ds^{+}=\frac{\hbar}{m}\frac{\nabla\psi^{+}}{\psi^{+}%
}ds^{+} \label{alpplus}%
\end{equation}
and
\begin{equation}
\alpha^{-}(X_{t},s^{-})ds^{-}=\frac{\hbar}{m}\frac{\nabla\psi^{-}}{\psi^{-}%
}ds^{-}, \label{alpminus}%
\end{equation}
and so are $X_{t}$ of Eq. (\ref{St1}).

Since the quantities in Eqs. (\ref{alpplus}) and (\ref{alpminus}) are mutually
complex conjugate, we may define%
\begin{equation}
\alpha^{+}(X_{t},s^{+})ds^{+}=\left[  \alpha^{\text{Real}}(X_{t}%
,t)+i\alpha^{\text{Imag}}(X_{t},t)\right]  dt
\end{equation}
and%
\begin{equation}
\alpha^{-}(X_{t},s^{-})ds^{-}=\left[  \alpha^{\text{Real}}(X_{t}%
,t)-i\alpha^{\text{Imag}}(X_{t},t)\right]  dt.
\end{equation}
$\alpha^{\text{Real}}(X_{t},t)$ and $\alpha^{\text{Imag}}(X_{t},t)$ are
readily obtained such that%

\begin{align}
\alpha^{\text{Real}}(X_{t},t)dt  &  =\frac{1}{2}\frac{\hbar}{m}\left(
\frac{\nabla\psi^{+}}{\psi^{+}}ds^{+}+\frac{\nabla\psi^{-}}{\psi^{-}}%
ds^{-}\right) \nonumber\\
&  =\frac{\hbar}{2m}\left(  \frac{\nabla\psi^{+}}{\psi^{+}}(-idt)+\frac
{\nabla\psi^{-}}{\psi^{-}}(idt)\right)
\end{align}
which gives rise to
\begin{align}
&  \alpha^{\text{Real}}(X_{t},t)dt\nonumber\\
&  =\frac{\hbar}{\rho m}dt\left(  \phi_{r}\nabla\phi_{c}-\phi_{c}\nabla
\phi_{r}\right) \nonumber\\
&  =-\frac{\hbar}{\rho m}\left(
\begin{array}
[c]{cc}%
\phi_{r} & \phi_{c}%
\end{array}
\right)  \text{\textbf{J}}\nabla\left(
\begin{array}
[c]{c}%
\phi_{r}\\
\phi_{c}%
\end{array}
\right)  dt \label{Realalpha}%
\end{align}
again with $\rho=\phi_{r}^{2}+\phi_{c}^{2}$. Recalling $\hat{p}=-$%
\textbf{J}$\hbar\nabla$, we see that the physical meaning of $\alpha
^{\text{Real}}$ turns out to be the locally normalized velocity at $X_{t}$ and
$t$ (see Eq. (\ref{v-density})). We also note that $\alpha^{\text{Real}}$ thus
attained is invariant with respect to any rotation of the vector $\left(
\begin{array}
[c]{cc}%
\phi_{r}(q,t) & \phi_{c}(q,t)
\end{array}
\right)  ^{T}$. \ After all we have%

\begin{align}
dX_{t}  &  =\alpha^{\text{Real}}(X_{t},t)dt+\sqrt{\frac{\hbar}{m}}%
dW_{0}\left(  t,\omega\right) \nonumber\\
&  =-\frac{\hbar}{\rho m}\left(
\begin{array}
[c]{cc}%
\phi_{r} & \phi_{c}%
\end{array}
\right)  \text{\textbf{J}}\nabla\left(
\begin{array}
[c]{c}%
\phi_{r}\\
\phi_{c}%
\end{array}
\right)  dt+\sqrt{\frac{\hbar}{m}}dW_{0}\left(  t,\omega\right) \nonumber\\
&  =-\frac{1}{\rho}\left(
\begin{array}
[c]{cc}%
\phi_{r} & \phi_{c}%
\end{array}
\right)  \hat{p}\left(
\begin{array}
[c]{c}%
\phi_{r}\\
\phi_{c}%
\end{array}
\right)  dt+\sqrt{\frac{\hbar}{m}}dW_{0}\left(  t,\omega\right)  ,
\label{Stchastic1}%
\end{align}
where the Wiener process is normalized such that%

\begin{equation}
dW_{0}\left(  t,\omega\right)  dW_{0}\left(  t,\omega\right)  =2Ddt=\frac
{\hbar}{m}dt \label{Dyndiff}%
\end{equation}
with
\begin{equation}
D=\frac{\hbar}{2m}. \label{Dconst}%
\end{equation}
Thus the stochastic path $X_{t}$ remains to run in the real-valued space.

Incidentally, the imaginary part of the velocity term turns out to be%

\begin{align}
&  \alpha^{\text{Imag}}(X_{t},t)dt=\frac{1}{2i}2D\left(  \frac{\nabla\psi^{+}%
}{\psi^{+}}ds^{+}-\frac{\nabla\psi^{-}}{\psi^{-}}ds^{-}\right) \nonumber\\
&  =-\frac{2D}{\rho}dt(\phi_{r}\left(  \nabla\phi_{r}\right)  +\phi_{c}\left(
\nabla\phi_{c}\right)  )\nonumber\\
&  =-\frac{D}{\rho}\left(  \nabla\rho\right)  dt.
\end{align}
We have currently no idea on the role of $\alpha^{\text{Imag}}(X_{t},t)$. \ \color{black}

\subsubsection{Triangle relation in the Schr\"{o}dinger dynamics}

\begin{figure*}[pth]
\centering\includegraphics[width=\textwidth]{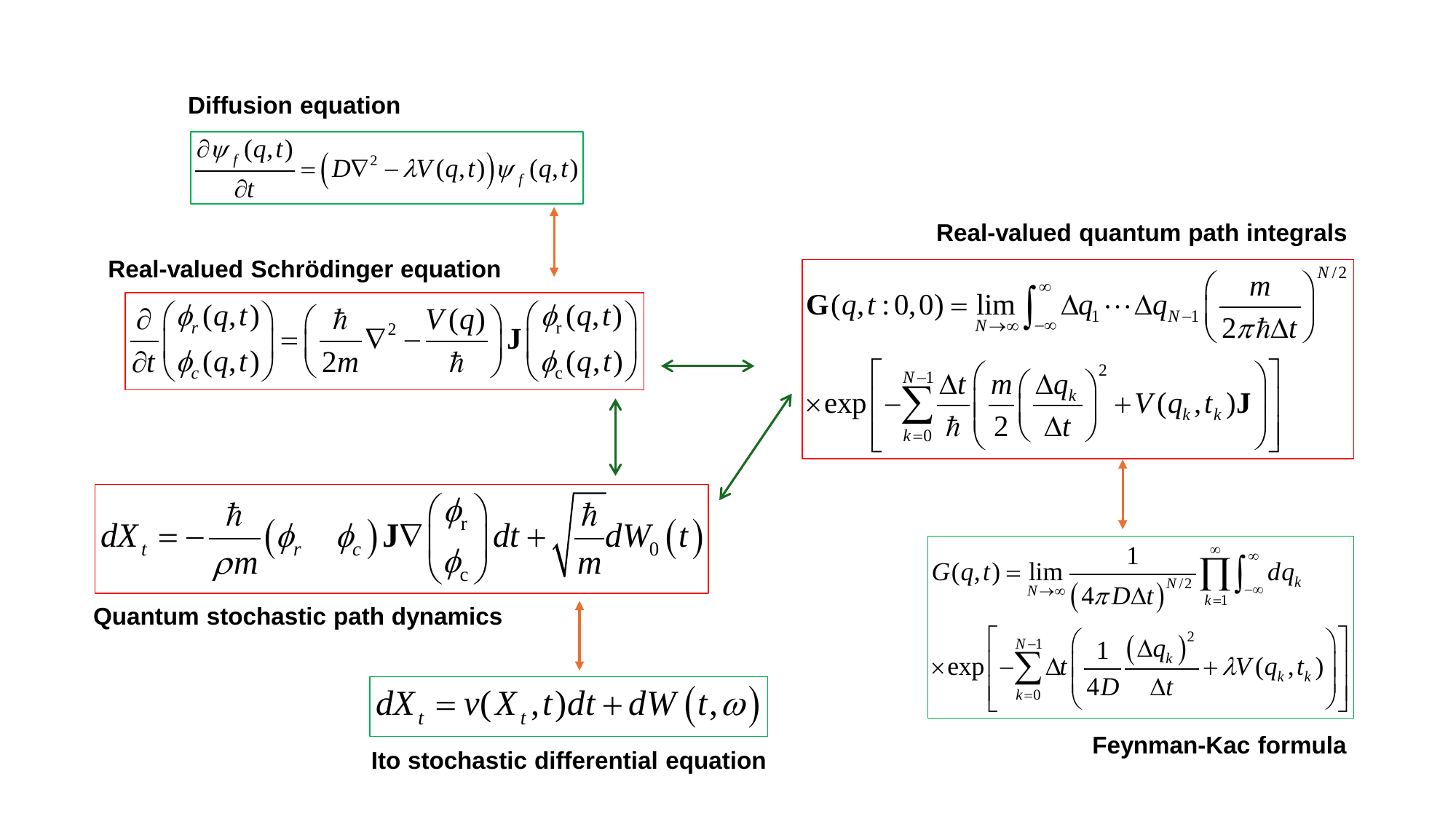}\caption{Triangle
relation in quantum dynamics similar to that of stochastic dynamics. }%
\label{Fig1}%
\end{figure*}

We have thus found the triangle theoretical structure in quantum dynamics;
Corresponding to the diffusion equation, the Feynman-Kac formula, and the
stochastic differential equation in stochastic dynamics, we have the
real-valued Schr\"{o}dinger equation Eq. (\ref{Sch45}), the real-valued
quantum path integrals Eq. (\ref{matrixGreen}), and the quantum stochastic
path dynamics Eq. (\ref{Stchastic1}). The real-valued Schr\"{o}dinger equation
is exactly equivalent to the original Schr\"{o}dinger equation, while the
real-valued quantum path integrals and quantum stochastic path dynamics have
been settled here. The triangle relation of the expressions to represent the
Schr\"{o}dinger dynamics along with the stochastic counterpart are diagrammed
in Fig. \ref{Fig1}.

\subsection{Time irreversibility}

The stochastic dynamics is time-irreversible in contrast to the
Schr\"{o}dinger equation, due to the presence of the Wiener process in it.
Yet, an average over the Wiener process leads to%

\begin{equation}
\left\langle \frac{dW_{0}}{dt}\right\rangle =0,
\end{equation}
which reduces the dynamics to%
\begin{equation}
\left\langle \frac{dX_{t}}{dt}\right\rangle =\left\langle \frac{1}{\rho
}\left(
\begin{array}
[c]{cc}%
\phi_{r} & \phi_{c}%
\end{array}
\right)  \frac{\hat{p}}{m}\left(
\begin{array}
[c]{c}%
\phi_{r}\\
\phi_{c}%
\end{array}
\right)  \right\rangle \label{avrX}%
\end{equation}
and therefore an average over the accumulated quantum paths turns out to be
time-reversal. This in turn indicates that one cannot single out a physical
path from the Schr\"{o}dinger equation directly or that the Schr\"{o}dinger
equation is not likely to give birth to time-irreversible paths directly by
definition. Also, it is rather questionable if the Nelson stochastic Newtonian
equation Eq. (\ref{Nelson}), which is time-reversal, can indeed describe the
quantum stochastic paths.

\subsection{Local velocity as a field for the quantum stochastic paths: Bohm
trajectory and Nagasawa path revisited}

Let us recall the de Broglie--Bohm theory claiming that the Schr\"{o}dinger
function serves as a pilot wave to guide particle
paths.\cite{auletta2001foundations} In the Bohm
representation,\cite{bohm1952suggested} the Schr\"{o}dinger function reads%
\begin{equation}
\psi\left(  q,t\right)  =R(q,t)\exp\left(  \frac{i}{\hbar}S_{B}\left(
q,t\right)  \right)  , \label{Bohm0}%
\end{equation}
and the quantum Hamilton-Jacobi (HJ) equation%
\begin{equation}
\frac{\partial S_{B}}{\partial t}+\frac{1}{2m}\left(  \nabla S_{B}\right)
^{2}+V-\frac{\hbar^{2}}{2m}\frac{\nabla^{2}R}{R}=0, \label{Bohm2}%
\end{equation}
is derived along with the equation of continuity for $R(q,t)^{2}$. If
$\psi\left(  q,t\right)  $ is given
beforehand,\cite{bohm1952suggested,Bohm-text,lopreore1999quantum,wyatt1999quantum,Wyatttext,sanz2013trajectory}
the quantum local velocity is given by\cite{de2016principles}%
\begin{equation}
v_{B}=\frac{1}{m}\nabla S_{B}=\frac{\hbar}{m}\operatorname{Im}\frac{\nabla
\psi}{\psi}. \label{Bohm3}%
\end{equation}
Therefore, $v_{B}$ is exactly the same as the drift velocity of Eq.
(\ref{Stchastic1}) at a common point $q$, provided that a common
Schr\"{o}dinger function is resorted to (see also Eq. (\ref{v-density})).
Hence, the presence of the Wiener process or not makes the mathematical
difference between the quantum paths and the Bohmian trajectories; a quantum
path wanders from one Bohmian trajectory to another in a stochastic manner to
the extent of $\sqrt{\hbar/m}$. The Bohmian trajectories represent a set of
integral curves of Eq. (\ref{Bohm2}). Sanz and Miret-Art{\'{e}}s describe that
each Bohmian trajectory represents the dynamics of a probe on the flow-lines
of $R(q,t)^{2}$ induced by $S_{B}$.\cite{sanz2013trajectory}

The quantum potential $-\hbar^{2}/2m(\nabla^{2}R/R)$ highlights the very
quantum nature involved in the Schr\"{o}dinger equation. Logically however it
seems hard for the Bohm representation to make an essentially novel
interpretation beyond the limit of the Schr\"{o}dinger equation, although it
has shed much new light on the hidden properties of the Schr\"{o}dinger
dynamics as in Eq. (\ref{Bohm3}%
).\cite{auletta2001foundations,home2007einstein,whitaker2012new,freire2022oxford}%

Instead of the de Broglie-Bohm pilot wave postulate, we may regard the local
velocity or the velocity drift term of Eq. (\ref{Stchastic1}) as a
\textquotedblleft guiding field in configuration space\textquotedblright\ on
which the paths to run (or be guided). In fact, we have already seen the
relation
\begin{equation}
\alpha^{\text{Real}}(X_{t},t)=\vec{v}^{\text{local}}(q,t)=\frac{\hbar}{m}%
\vec{\nabla}\theta(q,t)
\end{equation}
in the polar-coordinate representation of the real-valued Schr\"{o}dinger
vector as in Eq. (\ref{vangle}). (Note that $\theta(q,t)$ is not determined by
itself.) Besides, it holds $\vec{\nabla}\theta(q,t)=\vec{\nabla}S_{B}/\hbar$
at a common point $q$. Thus, in our language, the rotation of the
Schr\"{o}dinger vector\ is equipped implicitly as an intrinsic machinery to
drive the Schr\"{o}dinger dynamics. \color{black}

Wyatt and his colleague have figured out how to numerically integrate the
Bohmian paths without solving the Schr\"{o}dinger equation and have found
interesting applications.\cite{Wyatttext,sanz2013trajectory} Meanwhile, it is
already well known that the interference (fringe intensity) pattern in the
double-slit experiment is numerically realized by the set of the Bohmian
trajectories.\cite{philippidis1979quantum,sanz2013trajectory} Therefore a
sufficient large set of the present quantum stochastic paths should reproduce
the similar interference pattern (with stochastic fluctuation) in the path
distribution density, unless the Wiener process wipes away the pattern.

A precise study shows that the drift term of Eq. (\ref{Stchastic1}) and
Nagasawa's one in Eq. (\ref{Nagasawa}) are essentially the same as the
velocity drift term $\frac{\hbar}{m}\nabla(R+S)dt$ in the Ito scholastic
differential equation derived by Nagasawa, Eq. (\ref{Nagasawa}), even though
his stochastic equation has a correspondence with his own forward parabolic
differential equations, Eq. (\ref{Nagasawa1}), but not with the
Schr\"{o}dinger equation. Therefore the paths of Nagasawa and our quantum
stochastic paths presented here must mutually coincide for those of the same
initial conditions, provided that the Wiener processes happen to be the same
at each time. A notable aspect of the Nagasawa coupled equations, Eqs.
(\ref{Nagasawa1}) and (\ref{Nagasawa2}), and associated concepts may exceed
the Schr\"{o}dinger dynamics,\cite{nagasawa1989transformations,Nagasawa2}
while ours remains within the realm. Equation (\ref{Stchastic1}) seems rather
compact and more intuitively appealing. \ 

\section{The stochasticity as an intrinsic property of quantum dynamics
\label{Sec. Wiener}}

We here discuss the characteristic consequences of the stochasticity
manifesting in Eq. (\ref{Stchastic1}).

\subsection{Scaling law as a mathematical consequence of the Schr\"{o}dinger
dynamics}

In case where $\Delta q\rightarrow0$ and $\Delta t\rightarrow0$ cannot be
taken independently due to a constraint%
\begin{equation}
F(\Delta q,\Delta t)=0, \label{constraint}%
\end{equation}
it can follow that a derivative
\[
\lim_{\Delta t\rightarrow0\text{, }\Delta q\rightarrow0}\frac{\Delta q}{\Delta
t}%
\]
may not exist with $\Delta q\rightarrow0$. It is well known that the resultant
lack of the smoothness almost everywhere can give birth to a novel mathematics
like the stochastic calculus and their associated stochastic differential
equations by Ito and Stratonovich,\cite{oksendal2013stochastic} the fractal
geometry,\cite{mandelbrot1983fractal} and so on. The stochasticity in the
quantum path dynamics that appeared rather intuitively in Nelson's theory and
on the rigorous mathematical basis in the Nagasawa theory, is associated with
a scaling law as a specific case of Eq. (\ref{constraint}). This constraint is
mathematically of the same form as that of the Brownian motion, but it does
not mean that there exist random kickers surrounding a quantum particle. We
below study some consequences from the quantum Wiener process.

In the Feynman-Kac formula, Eq. (\ref{WienerG}), it is a usual practice that
the exponent is scaled such that
\begin{equation}
\left\langle \frac{1}{4D}\frac{\Delta q^{2}}{\Delta t}\right\rangle =\frac
{1}{2}, \label{randomB}%
\end{equation}
leading to the well-known expression $\left\langle \Delta q^{2}\right\rangle
=2D\Delta t$. This scaling law is applied to quantum dynamics, since exactly
the same scaling rule holds for the Wiener measure Eq. (\ref{WM}) in the
quantum mechanical extension of the Feynman-Kac formula Eq. (\ref{matrixGreen}%
). The exponent in the Feynman path integrals, Eq. (\ref{Fkernel}), can also
be scaled such that%

\begin{equation}
\left\langle \frac{m}{2\hbar}\frac{\Delta q^{2}}{\Delta t}i\right\rangle
=\frac{1}{2}i, \label{randomQF}%
\end{equation}
which anyway leads to
\begin{equation}
\left\langle \Delta q^{2}\right\rangle =\hbar/m\Delta t, \label{randomQ}%
\end{equation}
where $\left\langle \left(  q_{k+1}-q_{k}\right)  ^{2}\right\rangle
=\left\langle \Delta q^{2}\right\rangle $. Equation (\ref{randomQ}) is
consistent with $D=\hbar/2m$ of Eq. (\ref{Dconst}). Hence Eq. (\ref{randomQ})
tells that the two limiting processes $\Delta q\rightarrow0$ and $\Delta
t\rightarrow0$ cannot be taken independently.

We hence need to study an implication of the scaling relation $\left\langle
\Delta q^{2}\right\rangle =\hbar/m\Delta t$ or $\left(  dW_{0}\right)
^{2}=\left(  \hbar/m\right)  dt$ in quantum dynamics. Let us look back at Eq.
(\ref{randomQ}) in the form%
\begin{equation}
\left\langle \Delta q\times\left(  m\frac{\Delta q}{\Delta t}\right)
\right\rangle =\hbar. \label{uncert}%
\end{equation}
This expression implies that the convergence in the limit%

\begin{equation}
\lim_{\Delta t\rightarrow0,\Delta q\rightarrow0}\left\langle m\frac{\Delta
q}{\Delta t}\right\rangle =\bar{p},
\end{equation}
is not compatible with $\Delta q\rightarrow0$. Hence we must give up the
simultaneous determination of the momentum and the exact positioning (meaning
$\Delta q=0$) in the average sense. We also lose the conventional notion of
smoothness in $q(t)$ with respect to $t$.

\subsection{Hydrogen atom energy from the stochasticity}

To illustrate the physical significance of Eq. (\ref{uncert}), we exemplify a
one-dimensional hydrogen-like atom of nuclear charge $+Ze$. Equation
(\ref{uncert}) prohibits the electron from falling down to rest at the
position of the nucleus, because $\Delta q=0$ raises $\Delta q/\Delta t$ to
infinity. We then assume that the electron stays at a position of a distance
$\Delta q$ from the nucleus in average, and we may take $\Delta q/\Delta t$ as
a classical velocity (also in the average sense). The classical Hamiltonian%

\begin{equation}
H=\frac{1}{2}m\left(  \frac{\Delta q}{\Delta t}\right)  ^{2}-\frac{Ze^{2}%
}{\Delta q}.
\end{equation}
is then constrained by Eq. (\ref{uncert}), that is,%
\begin{equation}
\frac{\Delta q}{\Delta t}=\frac{\hbar}{m}\frac{1}{\Delta q},
\end{equation}
and is reduced to
\begin{equation}
H=\frac{1}{2}\frac{\hbar^{2}}{m}\Delta q^{-2}-Ze^{2}\Delta q^{-1}.
\end{equation}
The condition of force balance
\begin{equation}
\frac{dH}{d(\Delta q)}=-\frac{\hbar^{2}}{m}\Delta q^{-3}+Ze^{2}\Delta
q^{-2}=0,
\end{equation}
finds the solutions $\left\vert \Delta q\right\vert =\infty$ and%
\begin{equation}
\Delta q=\frac{\hbar^{2}}{Zme^{2}}=\frac{a_{0}}{Z}. \label{hAtom}%
\end{equation}
This last solution makes the lowest energy state. It is noticed that this
$\Delta q$ is exactly the same\ as the Bohr radius $a_{0}$, the radius of the
ground state hydrogen-like atom in the Bohr model.\cite{Messiah} The energy
$E$ follows to be
\begin{equation}
E=-\frac{1}{2}\frac{Z^{2}me^{4}}{\hbar^{2}}, \label{BohrE}%
\end{equation}
which is the ground-state energy of the hydrogen-like atom.\cite{Messiah} We
have adopted neither the the notion of the wavelength of matter wave nor the
force balance between the centrifugal force and Coulombic
attraction.\cite{Messiah}

Further, one can consider the effect of harmonics of those oscillatory
integrals by modifying Eq. (\ref{randomQ}) in the Feynman path integrals Eq.
(\ref{Fkernel}) such that
\begin{equation}
\left\langle \frac{m}{2\hbar}\frac{\Delta q^{2}}{\Delta t}\right\rangle
i=\frac{1}{2}ni,
\end{equation}
with $n$ being an integer. This is just
\begin{equation}
\left\langle m(\frac{\Delta q}{\Delta t})\Delta q\right\rangle =n\hbar,
\label{BohrQ}%
\end{equation}
leading to the Bohr quantization condition for hydrogen-like atom%
\begin{equation}
mvr=n\hbar,
\end{equation}
where $r$ is the radius of electron motion around the nucleus.\cite{Messiah}
By replacing $\hbar$ with $n\hbar$ in the quantities like Eq. (\ref{hAtom})
and (\ref{BohrE}), we can extend it to the excited states. (For the canonical
semiclassical quantization for integrable
system\cite{einstein1917quantensatz,Messiah,Brack} and for nonintegrable
and/or chaotic
systems.\cite{Gutzwiller1,Gutzwiller2,kleinert2006path,Reichl,Gaspard1995,RobnikQuantumChaos,KTChaos}%
)

The above primitive example suggests a deep relationship between quantum
dynamics and stochastic dynamics and reminds of the comment from Einstein to
Heisenberg, \textquotedblleft You can see and track the orbit of an electron
in the cloud chamber. Nevertheless, you intend to entirely deny the notion of
orbit in an atom, don't you?\textquotedblright\ and Heisenberg replies,
\textquotedblleft We cannot observe any orbit of an electron in an atom.
\ldots\ Only the observable quantities should be treated by a
theory.\textquotedblright\ (W. Heisenberg in \textquotedblleft Der teil und
das ganze\textquotedblright) Note again that the condition of Eq.
(\ref{uncert}) has not been brought in by an external noise. Also, it is not
merely a matter of scaling or a simple analogy to the Brownian motion. We
regard the Wiener process in the dynamics of the quantum path Eq.
(\ref{Stchastic1}) as a manifestation of the essential quantum nature, which
is smoothed away from the Schr\"{o}dinger equation.

\subsection{Uncertainty relations from the
stochasticity\label{sec:uncertainty}}

Since we have no random kickers surrounding the quantum particles, it does not
make sense to interpret $\Delta q$ as a displacement by an impulse. A
classical particle is supposed to have a definite position and momentum as an
intrinsic property inherent to each one. They can be determined by tuning the
time interval to an arbitrary length. However, the quantum scaling law does
not allow for such a unique determination. Suppose a situation of $V=0$ and
average velocity $\bar{v}=0$ at a given point. Then we may regard $\Delta q$
as an average length of an area, in the outside of which the particle is not
observed during the interval $\Delta t$. In an attempt to determine $\Delta q$
precisely, one may want to make $\Delta t$ shorter aiming at $\Delta
q\rightarrow0$. However, $\Delta q$ can become smaller only \textquotedblleft
more slowly\textquotedblright\ according to Eq. (\ref{randomQ}) in such a
manner that making $\Delta q\rightarrow\Delta q/2$ requires $\Delta
t\rightarrow\Delta t/4$. The shutter speed of camera must be made faster by 4
times to capture it in a two times narrower space. Therefore, symbolically,
quantum particle cannot be located at a given point by an operation of $\Delta
t\rightarrow0$.\color{black}

It is therefore quite natural to expect that the present quantum stochasticity
should brings about an uncertainty relation between the relevant
quantities.\cite{Messiah,ozawa2003universally,freire2022oxford} First, we
suppose a particle residing at $\left(  q_{0},p_{0}\right)  $ in classical
phase space. Consider a width $\Delta q$ in configuration space $q$. In the
expression of Eq. (\ref{uncert}), we may conceive that the stochastic dynamics
can induce an additional momentum $m\Delta q/\Delta t$ in this interval
$\Delta q$. And, this additional momentum, say, $\Delta p$ gives an
uncertainty to the momentum to $p_{0}$, or, $p_{0}\rightarrow$ $p_{0}+\Delta
p$. In this sense, Eq. (\ref{uncert}) may read
\begin{equation}
\left\langle \Delta p\Delta q\right\rangle =\hbar. \label{qpdelta}%
\end{equation}
Since the particle may make a zigzag motion before getting out of the space
interval $\Delta q$, $\Delta p$ can be larger, and therefore we expect
\begin{equation}
\left\langle \Delta p\Delta q\right\rangle \geq\hbar. \label{uncertainty2}%
\end{equation}

The above description may cause a misunderstanding that a shorter space range
$\Delta q$ for a given time interval $\Delta t$ should give a smaller $\Delta
p$ against Eq. (\ref{qpdelta}). However, the reality is that if $\Delta q$ is
scaled such that $\Delta q\rightarrow\Delta q/N$, the stochasticity relation
scales the time $\Delta t\rightarrow\Delta t/N^{2}$ and $\Delta p\rightarrow
N\Delta p$. Therefore small range $\Delta q$ makes a larger stochastic
momentum $\Delta p$. The uncertainty Eq. (\ref{uncertainty2}) claims that the
stochasticity prevents specifying precise information in the cell smaller than
the size $\left\langle \Delta p\Delta q\right\rangle =\hbar$, which is already
a common sense in quantum mechanics.

Likewise we consider $\Delta q$ and rewrite the stochastic relation as%
\begin{equation}
\left\langle \frac{m}{2}(\frac{\Delta q}{\Delta t})^{2}\Delta t\right\rangle
=\frac{\hbar}{2}.
\end{equation}
The stochasticity may bring about an additional kinetic energy $m(\Delta
q/\Delta t)^{2}/2=\Delta E_{K}$, and $\Delta E_{K}$ is subject to%
\begin{equation}
\left\langle \Delta E_{K}\Delta t\right\rangle =\frac{\hbar}{2}.
\end{equation}
Since the particle can move in a zigzag motion, the stochastically induced
kinetic energy can be larger than this $\Delta E_{K}$, and we have
\begin{equation}
\left\langle \Delta E_{K}\Delta t\right\rangle \geq\frac{\hbar}{2}.
\label{uncertainty3}%
\end{equation}
Again we note that if $\Delta t\rightarrow\Delta t/N$, then $\Delta
q\rightarrow\Delta q/\sqrt{N}$ and $\Delta E_{K}\rightarrow N\Delta E_{K}$.
Thus the shorter time interval induces a larger stochastically uncertain
kinetic energy.

We note that the present uncertainty has arisen in a manner different from the
Heisenberg's one, which reflects the operational disturbance inevitably
introduced to an observation process, from the
Kennard-Robertson-Schr\"{o}dinger principles, which originate from the
property of the Schr\"{o}dinger function such as the universal relationship
between a configuration-space distribution function and its Fourier
transform,\cite{dirac1981principles,Messiah,auletta2001foundations,freire2022oxford}
and from the Ozawa's principle deeply unifying
them.\cite{ozawa2003universally,ozawa2015heisenberg} Meanwhile, the present
uncertainty comes from the quantum property that is not present in the
Schr\"{o}dinger equation.

\section{Indirect correlation among the quantum paths \label{sec:correlation}}

We next study the properties related mainly to the velocity drift term in the
quantum stochastic path dynamics.

\subsection{Quantum canonical equations of motion}

To better understand the quantum stochastic path dynamics, we transform it to
the form of the quantum canonical equations of motion. First, the
time-variation of the unnormalized momentum term $\left(
\begin{array}
[c]{cc}%
\phi_{r} & \phi_{c}%
\end{array}
\right)  \hat{p}\left(
\begin{array}
[c]{cc}%
\phi_{r} & \phi_{c}%
\end{array}
\right)  ^{T}$ in Eq. (\ref{Stchastic1}) should be subjected to the Heisenberg
equation of motion of Eq. (\ref{Heisenberg}) such that%

\begin{align}
&  \frac{d}{dt}\left(
\begin{array}
[c]{cc}%
\phi_{r} & \phi_{c}%
\end{array}
\right)  \hat{p}\left(
\begin{array}
[c]{c}%
\phi_{r}\\
\phi_{c}%
\end{array}
\right)  =\frac{1}{\hbar}\left(
\begin{array}
[c]{cc}%
\phi_{r} & \phi_{c}%
\end{array}
\right)  \left[  \hat{H},\hat{p}\right]  \text{\textbf{J}}\left(
\begin{array}
[c]{c}%
\phi_{r}\\
\phi_{c}%
\end{array}
\right) \nonumber\\
&  =-\left(
\begin{array}
[c]{cc}%
\phi_{r} & \phi_{c}%
\end{array}
\right)  \left[  V,\text{\textbf{J}}\vec{\nabla}\right]  \text{\textbf{J}%
}\left(
\begin{array}
[c]{c}%
\phi_{r}\\
\phi_{c}%
\end{array}
\right) \nonumber\\
&  =\left(
\begin{array}
[c]{cc}%
\phi_{r} & \phi_{c}%
\end{array}
\right)  \left(  -\vec{\nabla}V\right)  \left(
\begin{array}
[c]{c}%
\phi_{r}\\
\phi_{c}%
\end{array}
\right) \nonumber\\
&  =\left(  \vec{\nabla}V\right)  \rho. \label{Ehrenfestp}%
\end{align}
This representation is an alternative expression of the Ehrenfest
theorem,\cite{Messiah} although the integration over the $q$-coordinates is
not performed. We may thus formally combine Eqs. (\ref{Stchastic1}) and
(\ref{Ehrenfestp}) into a set
\begin{equation}
\left\{
\begin{array}
[c]{c}%
dX_{t}=\frac{1}{m\rho}P_{X_{t}}dt+\sqrt{\frac{\hbar}{m}}dW_{0}\left(
t,\omega\right) \\
\\
dP_{X_{t}}=-\left(  \vec{\nabla}V\right)  \rho dt\text{
\ \ \ \ \ \ \ \ \ \ \ \ \ \ \ }%
\end{array}
\right.  \label{HM12}%
\end{equation}
with the explicit definition of the unnormalized local momentum%
\begin{equation}
P_{X_{t}}=\left.  \left(
\begin{array}
[c]{cc}%
\phi_{r} & \phi_{c}%
\end{array}
\right)  \hat{p}\left(
\begin{array}
[c]{c}%
\phi_{r}\\
\phi_{c}%
\end{array}
\right)  \right\vert _{X_{t}}. \label{PXlarge}%
\end{equation}
\color{black}

\subsection{Quantum Newtonian equation}

The coupled equation in Eq. (\ref{HM12}) is further combined into a one piece
of expression, that is%

\begin{align}
d^{2}X_{t}  &  =\frac{1}{m\rho}dP_{X_{t}}dt-\frac{P_{X_{t}}}{m\rho^{2}}d\rho
dt+\sqrt{\frac{\hbar}{m}}d^{2}W_{0}\left(  t,\omega\right) \nonumber\\
&  =-\frac{1}{m}\vec{\nabla}V\left(  dt\right)  ^{2}-\frac{1}{\rho}\vec
{v}^{\text{local}}d\rho dt+\sqrt{\frac{\hbar}{m}}d^{2}W_{0}\left(
t,\omega\right)  , \label{QNewton1}%
\end{align}
which we may refer to as the quantum Newtonian equation. Notice that this
expression is totally different from the Nelson's postulated equation of Eq.
(\ref{Nelson}).

To proceed further, we recall Eq. (\ref{primeflux}) and Eq. (\ref{v-density})
such that%

\begin{align}
\frac{\partial\rho}{\partial t}  &  =-\vec{\nabla}\cdot\vec{j}=-\vec{\nabla
}\cdot\frac{\hbar}{m}\left(  \phi_{r}\vec{\nabla}\phi_{c}-\phi_{c}\vec{\nabla
}\phi_{r}\right) \nonumber\\
&  =-\vec{\nabla}\cdot(\rho\vec{v}^{\text{local}}(q,t))
\end{align}
giving rise to%

\begin{equation}
d\rho=-\vec{\nabla}\cdot(\rho\vec{v}^{\text{local}}(q,t))dt,
\end{equation}
and therefore the quantum Newtonian equation is rewritten as%

\begin{align}
d^{2}X_{t}  &  =-\frac{1}{m}\vec{\nabla}V\left(  dt\right)  ^{2}-\frac{1}%
{\rho}\vec{v}^{\text{local}}d\rho dt+\sqrt{\frac{\hbar}{m}}d^{2}W_{0}\left(
t,\omega\right) \nonumber\\
&  =-\frac{1}{m}\vec{\nabla}V\left(  dt\right)  ^{2}+\frac{1}{\rho}\vec
{v}^{\text{local}}\left(  \vec{\nabla}\cdot(\rho\vec{v}^{\text{local}%
})\right)  \left(  dt\right)  ^{2}\nonumber\\
&  +\sqrt{\frac{\hbar}{m}}d^{2}W_{0}\left(  t,\omega\right)  \label{QNewton2}%
\end{align}
It is clear that the genuine quantum terms are the last two terms in this expression.

\subsection{Classical limit of the quantum stochastic motion}

\subsubsection{To the Hamilton canonical equations of motion and Newtonian
equation}

Let us consider the classical limit of the quantum canonical equations of
motion of Eq. (\ref{HM12}) by setting $\hbar\rightarrow0$. The Wiener process
is reduced to zero, simply because it is linear in $\sqrt{\hbar}$. Next,
rewrite Eq. (\ref{PXlarge}) in the polar coordinate as in Eq. (\ref{qdist}) to find%

\begin{equation}
P_{X_{t}}=\hbar\rho\nabla\theta(X_{t},t). \label{PX}%
\end{equation}
Because the absence of the Wiener process in the classical limit, a quantum
path does not have a chance of branching in its direction and is smooth
(differentiable). Therefore $X_{t}$ is uniquely tied to the initial position
$X_{0}$. Therefore, the magnitude at $\rho(X_{0},0)$ should remain constant
along the path such that
\begin{equation}
\rho(X_{t},t)=\rho(X_{0},0)\equiv\rho_{0}, \label{ClassicalC}%
\end{equation}
modifying $P_{X_{t}}$ of Eq. (\ref{PX}) to
\begin{equation}
P_{X_{t}}=\hbar\rho_{0}\nabla\theta(X_{t},t).
\end{equation}
(Yet, $\hbar\rightarrow0$ should not be taken in this expression, because
$\nabla\theta(X_{t},t)$ gives birth to a term proportional to $\hbar^{-1}$.)
Further, we may define the classical momentum $p_{cl}(X_{t},t)$ as%

\begin{equation}
p_{cl}(X_{t},t)\equiv\lim_{\hbar\rightarrow0}\hbar\nabla\theta(X_{t},t),
\label{thetap}%
\end{equation}
which brings about
\begin{equation}
P_{X_{t}}=\hbar\rho_{0}\nabla\theta(X_{t},t)\rightarrow\rho_{0}p_{cl}(X_{t},t)
\label{Pp}%
\end{equation}
and%
\begin{equation}
dP_{X_{t}}\rightarrow\rho_{0}dp_{cl}(X_{t},t). \label{Pp2}%
\end{equation}
Inserting Eq. (\ref{Pp}) into the first equation of Eq. (\ref{HM12}), we have
\begin{equation}
dX_{t}=\frac{1}{m}p_{cl}(X_{t},t)dt. \label{CLH1}%
\end{equation}
Likewise, Eq. (\ref{Pp2}) revises Eq. (\ref{HM12}) to%

\begin{equation}
dP_{X_{t}}=-\rho_{0}\vec{\nabla}V\left(  X_{t}\right)  dt. \label{Vclassical}%
\end{equation}
Then the combination of Eqs. (\ref{Pp2}) and (\ref{Vclassical}) results in%
\begin{equation}
dp_{cl}(X_{t},t)=-\vec{\nabla}V\left(  X_{t}\right)  dt. \label{CLH2}%
\end{equation}
Because there is no stochastic term in these expressions, we can take the
simple limit $dX_{t}\rightarrow0$ and $dp_{cl}(X_{t},t)\rightarrow0$ as
$dt\rightarrow0$, and thereby the Hamilton canonical equations of motion follows%

\begin{equation}
\left\{
\begin{array}
[c]{c}%
\frac{dX_{t}}{dt}=\frac{1}{m}p_{cl}(X_{t},t)\text{ \ \ }\\
\\
\frac{dp_{cl}(X_{t},t)}{dt}=-\vec{\nabla}V\left(  X_{t}\right)  .
\end{array}
\right.  \label{Canonical}%
\end{equation}

The classical limit of the quantum Newtonian equation is taken in Eq.
(\ref{QNewton2}) by putting
\begin{equation}
d\rho=0
\end{equation}
along a trajectory, and the zero Wiener process. The result turns out to be
\begin{equation}
d^{2}X_{t}=-\frac{1}{m}\vec{\nabla}V\left(  dt\right)  ^{2}.
\end{equation}

The WKB theory\cite{Schiff, Messiah} and the Bohm representation in Eq.
(\ref{Bohm2}) as well show the clear correspondence between the
Schr\"{o}dinger equation and the Hamilton-Jacobi equation. The classical limit
of the equation of motion for the Wigner phase-space distribution function is
the classical Liouville
equation.\cite{wigner1932quantum,zachos2005quantum,polkovnikov2010phase} These
are primarily the classical limit for the distribution functions. On the other
hand, the quantum path dynamics turns out to be reduced directly to the
classical path dynamics like the Hamilton canonical equations of motion. This
is one of the reflections of the dual structure of the Schr\"{o}dinger dynamics.

\subsubsection{Insights from taking the classical limits}

The above study on the classical limit of the quantum path dynamics highlights
the intrinsic quantum effects as follows.

1. Here again, the classical limit suggests the critical role of the Wiener
process in quantum path dynamics. We have observed that \textquotedblleft if a
quantum path could be tracked in a deterministic manner due to the
disappearance of the quantum stochasticity, it should be subjected to the law
of classical mechanics.\textquotedblright\ Then, the contraposition of this
statement is that \textquotedblleft if a quantum path does not satisfy
classical mechanics, it cannot be tracked in a deterministic manner because of
the presence of stochasticity.\textquotedblright\ 

2. The quantum effects are expected to appear more significantly when the
condition of Eq. (\ref{ClassicalC}) is violated and $\rho(X_{t},t)$ has a
broader spatial distribution around $X_{t}$, which indicates that the
stochastic paths can wander about the area covered by $\left\vert
\psi(q,t)\right\vert ^{2}$.\cite{Paper-II}

3. The time derivative of the velocity drift term depends on $\rho(X_{t},t)$
in the form $\rho(X_{t},t)\vec{\nabla}V(X_{t},t)$ as in Eq. (\ref{HM12}),
whereas the quantum paths contribute to the formation of $\rho(X_{t},t)$.
Therefore, it is confirmed that there is a self-referential nonlinear relation
between the parts (the quantum path) and whole ($\rho(X_{t},t)$ and
$\psi(q,t)$) in the full quantum dynamics. In the classical limit, such a
nonlinear dynamical relation is dissolved.

\subsection{Indirect correlation \label{Sec:correlation}}

\subsubsection{Correlation among the quantum paths}

Each quantum stochastic path $X_{t}(\omega)$ does not interfere with others on
the way of propagation, and each represents independent event. (In contrast, a
single Schr\"{o}dinger function can generally represent a coherent
superposition of many possible events.) This is similar to the Brownian
dynamics in which each path represents an individual sample of a single
Brownian motion in real space. An infinite number of possible Brownian paths
are theoretically supposed to run on different occasions in a manner one after
another one. (We disregard collision among the Brownian particles.) Therefore,
no direct interaction among the stochastic paths is expected, in contrast to
the coherent interaction among those complex paths in the Feynman integrals.

Nevertheless, a quantum path can \textquotedblleft
indirectly\textquotedblright\ correlate with other quantum paths through the
velocity drift term $\alpha^{\text{Real}}(X_{t},t)$ in Eq. (\ref{Stchastic1}),
which is a functional of the Schr\"{o}dinger function. We mean by indirect
correlation that (i) there is no machinery through which two or more quantum
paths, launched on different occasions, mechanically interact each other, but
(ii) as long as a same Schr\"{o}dinger function lies behind the velocity drift
velocity of Eq. (\ref{Stchastic1}), not only they conform to the same global
trend but also each is mathematically responsible to reproduce the
$\alpha^{\text{Real}}(X_{t},t)$ as a whole through the Schr\"{o}dinger
function. Therefore, even a slight change of the experimental context in the
level of the quantum path can affect the Schr\"{o}dinger function, and thereby
it can modulate the way of correlation among the paths. The quantum paths are
thus supposed to run under the mutual indirect correlation. We will be back to
this aspect later.\color{black}

\subsubsection{Interference pattern in the double slit experiment}

Each quantum path should be able to pass through only one of the two slits in
the double-slit experiment, because a single quantum path does not branch. On
the other hand, a single Schr\"{o}dinger function can bifurcate and pass
through the two slits simultaneously as a coherent distribution function. Note
that it is not $\left\vert \psi(q,t)\right\vert ^{2}$ that physically makes
individual spots on the measurement board, but each quantum path does one by
one. Nevertheless, each quantum path is \textquotedblleft driven and
guided\textquotedblright\ by the drift velocity term, which is composed of the
relevant Schr\"{o}dinger function $\psi(q,t)$, and the quantum paths take the
mathematically same routes as the Bohmian trajectories if the Wiener process
is ignored. The density distribution of the Bohmian trajectories are well
known to reproduce the interference
pattern.\cite{philippidis1979quantum,sanz2013trajectory} It is therefore not
very mysterious that the interference pattern is shaped after many launchings
of single particle. It is a great mystery, however, how nature manages to
materialize the nonlinear relation between the parts (quantum paths) and the
whole (the Schr\"{o}dinger function).\ 

\color{black}

\subsection{Quantum entanglement manifesting on a\ single quantum path and
spontaneous detanglement}

\subsubsection{Persisting entanglement}

It could be doubted whether the quantum entanglement can be fully taken into
account by a single quantum stochastic path. This question is rather natural
because the quantum entanglement is considered a superposition of plural local
states to be persisted in asymptotic regions.

Suppose as an example a very simple two-particle Schr\"{o}dinger function
\begin{equation}
\psi(q_{1}^{\alpha},q_{2}^{\beta},t)=N(t)\left(  a(q_{1}^{\alpha}%
)b(q_{2}^{\beta})-b(q_{1}^{\alpha})a(q_{2}^{\beta})\right)  \label{ent}%
\end{equation}
in which two local functions $\left\{  a(q_{1}^{\alpha}),b(q_{2}^{\beta
})\right\}  $ are entangled, where $q_{1}^{\alpha}$ ($q_{2}^{\beta}$) is the
short-hand notation of spatial and spin coordinates. The density has the
following permutation symmetry
\begin{equation}
\rho(q_{1}^{\alpha},q_{2}^{\beta},t)=\rho(q_{2}^{\beta},q_{1}^{\alpha},t).
\end{equation}
The quantum stochastic path dynamics for the electronic coordinates $\left(
X_{1}^{\alpha},X_{2}^{\beta}\right)  $ reads in this case%
\begin{align}
X_{1}^{\alpha}(t+dt) &  =X_{1}^{\alpha}(t)+\frac{1}{\rho m}P_{X_{t}}%
^{(1)}(X_{1}^{\alpha}(t),X_{2}^{\beta}(t))dt\nonumber\\
&  +\sqrt{\frac{\hbar}{m}}dW_{0}^{\left(  1\right)  }\left(  t,\omega\right)
\label{X1}%
\end{align}
and%
\begin{align}
X_{2}^{\beta}(t+dt) &  =X_{2}^{\beta}(t)+\frac{1}{\rho m}P_{X_{t}}^{(2)}%
(X_{1}^{\alpha}(t),X_{2}^{\beta}(t))dt\nonumber\\
&  +\sqrt{\frac{\hbar}{m}}dW_{0}^{\left(  1\right)  }\left(  t,\omega\right)
,\label{X2}%
\end{align}
where%
\begin{align}
&  P_{X_{t}}^{(i)}(X_{1}^{\alpha},X_{2}^{\beta},t)\nonumber\\
&  =\left(
\begin{array}
[c]{cc}%
\operatorname{Re}(\psi(X_{1}^{\alpha},X_{2}^{\beta},t)) & \operatorname{Im}%
(\psi(X_{1}^{\alpha},X_{2}^{\beta},t))
\end{array}
\right)  \nonumber\\
\times &  \hat{p}_{i}\left(
\begin{array}
[c]{c}%
\operatorname{Re}(\psi(X_{1}^{\alpha},X_{2}^{\beta},t))\\
\operatorname{Im}(\psi(X_{1}^{\alpha},X_{2}^{\beta},t))
\end{array}
\right)
\end{align}
with $i=1,2$. It is obvious the two velocity drift terms are symmetric with
respect to the permutation between the positions $X_{1}^{\alpha}$ and
$X_{2}^{\beta}$. The Wiener processes $dW_{0}^{\left(  1\right)  }\left(
t,\omega\right)  $ and $dW_{0}^{\left(  2\right)  }\left(  t,\omega\right)  $
take place independently for each path (meaning $dW_{0}^{\left(  1\right)
}\left(  t,\omega\right)  \neq dW_{0}^{\left(  2\right)  }\left(
t,\omega\right)  $) and do not depend on the electron spin. Likewise, the
force terms in Eq. (\ref{HM12}) working on the the coordinates $X_{1}^{\alpha
}$ and $X_{2}^{\beta}$ are%
\begin{align}
&  \frac{d\vec{P}_{X_{t}}^{(i)}}{dt}=-\rho(X_{1}^{\alpha},X_{2}^{\beta}%
,t)\vec{\nabla}_{i}V(X_{1}^{\alpha},X_{2}^{\beta})\nonumber\\
&  =-\vec{\nabla}_{i}(V(X_{1}^{\alpha},X_{2}^{\beta})\nonumber\\
&  \times\left\vert N_{\pm}(t)\right\vert ^{2}[\left\vert a(X_{1}^{\alpha
})\right\vert ^{2}\left\vert b(X_{2}^{\beta})\right\vert ^{2}+\left\vert
b(X_{1}^{\alpha})\right\vert ^{2}\left\vert a(X_{2}^{\beta})\right\vert
^{2}\nonumber\\
&  +\left(  a^{\ast}(X_{1}^{\alpha})b(X_{1}^{\alpha})b^{\ast}(X_{2}^{\beta
})a(X_{2}^{\beta})+\text{c.c.}\right)  ],
\end{align}
($i=1,2$) and again they are symmetric with respect to the permutation between
$i=1$ and $i=2$. Thus the present two electron system feels a spin-free
entangled force, and therefore, if the Wiener processes are ignored,
$X_{1}^{\alpha}$ and $X_{2}^{\beta}$ keep symmetric as the Schr\"{o}dinger
function in Eq. (\ref{ent}) does.

On the other hand, the Wiener process keeps randomizing the deterministic
process including the above entanglement,\ up to an extent proportional to
$\left(  \hbar/m\right)  ^{1/2}$. Nevertheless, at each time step $dt$,
$dP_{X_{t}}$ is updated under the pure entanglement force with no quantum
stochasticity. Therefore, the quantum path $X_{t}$ is supposed to mostly
survive the randomization and maintain the effect from the entanglement.

\subsubsection{Detanglement}

A subtle balance between the entanglement and the stochastic dynamics can
begin when the orbital overlap between $a(q)$ and $b(q)$ becomes small, that
is, $a(q)b(q)\sim0$, and the entanglement eventually comes to end. This is
because the energy gap between the states of $a(q_{1}^{\alpha})b(q_{2}^{\beta
})-b(q_{1}^{\alpha})a(q_{2}^{\beta})$, $a(q_{1}^{\alpha})b(q_{2}^{\beta})$,
and $b(q_{1}^{\alpha})a(q_{2}^{\beta})$ become small and can mix in a way to
violate the permutation symmetry.

We recall the energy-time uncertainty due to the quantum stochasticity (see
Sec. \ref{sec:uncertainty}). Suppose that the stochasticity due to the Wiener
process $\sqrt{\frac{\hbar}{m}}dW_{0}\left(  t,\omega\right)  $ in Eq.
(\ref{Stchastic1}) gives birth to a small shift of the position $\Delta X_{t}%
$. Then, Eq. (\ref{uncertainty2}) can gives rise to the uncertainty in
momentum according to $\left\langle \Delta p\Delta q\right\rangle \geq\hbar$
and a time displacement $\Delta t$ to integrate Eq. (\ref{Stchastic1}) should
bring about the uncertain width in the kinetic energy due to $\left\langle
\Delta E_{K}\Delta t\right\rangle \geq\hbar/2$ of Eq. (\ref{uncertainty3}).
Therefore, if there exist additional velocity drift fields nearby in energy,
the path can jump to one of them. Therefore, the symmetry breaking due to the
stochasticity in ($X_{1}^{\alpha}$,$X_{2}^{\beta}$)-space can effectively
makes it possible for the quantum path to jump to one of the velocity fields
made up by $a(q_{1}^{\alpha})b(q_{2}^{\beta})$ and $b(q_{1}^{\alpha}%
)a(q_{2}^{\beta})$ .

On the other hand, even after the particles are separated far from each other
to asymptotic areas, say, $q($A$)$ and $q($B$)$ with $\left\vert
q(\text{A})-q(\text{B})\right\vert \sim\infty$ and thereby $a(q($%
A$))b(q($B$))=0$, the Schr\"{o}dinger equation alone has no mechanism to
decouple the entanglement, since this is essentially a matter of symmetry. The
asymptotic Schr\"{o}dinger function thus ends up with the form%
\begin{align}
&  \psi^{asymptotic}(q_{1}^{\alpha},q_{2}^{\beta},t)\nonumber\\
&  =N(t)\left(  a^{A}(q_{1}^{\alpha}(\text{A}))b^{B}(q_{2}^{\beta}%
(\text{B}))-b^{B}(q_{1}^{\alpha}(\text{B}))a^{A}(q_{2}^{\beta}(\text{A}%
))\right)  , \label{entasymp}%
\end{align}
where $a^{A}$ and $b^{B}$ are the asymptotic functions of $a$ and $b$ in the
$q($A$)$ and $q($B$)$ areas, respectively. In integrating the quantum
stochastic path dynamics, however, only one of the two corresponding channels%
\begin{equation}
(X_{1}^{\alpha}(\text{A}),X_{2}^{\beta}(\text{B})) \label{X3}%
\end{equation}
and%
\begin{equation}
(X_{2}^{\beta}(\text{A}),X_{1}^{\alpha}(\text{B})), \label{X4}%
\end{equation}
can be realized asymptotically, where $X_{1}^{\alpha}($A$)$ indicates
$X_{1}^{\alpha}$ found in the $q($A$)$ area. The paths of Eq. (\ref{X3}) and
(\ref{X4}) are not subject to the permutation symmetry with each other. Only
one of them can happen in a single event, and it is with the same probability
for these two to take place in a large ensemble of the relevant experiment.
Thus the quantum entanglement can be resolved spontaneously in an
uncontrollable manner. \color{black}

Here we notice that the statement \textquotedblleft the channel selection
$(X_{1}^{\alpha}($A$),X_{2}^{\beta}($B$))$ or $(X_{2}^{\beta}($A$),X_{1}%
^{\alpha}($B$))$ is \textit{stochastically} determined by the quantum
spontaneous Wiener process\textquotedblright\ can be effectively replaced with
a statement \textquotedblleft the channel selection $\left\vert a^{A}%
(X_{1}^{\alpha}(\text{A}))\right\vert ^{2}\left\vert b^{B}(X_{2}^{\beta
}(\text{B}))\right\vert ^{2}$ or $\left\vert b^{B}(X_{1}^{\alpha}%
(\text{B}))\right\vert ^{2}\left\vert a^{A}(X_{2}^{\beta}(\text{A}%
))\right\vert ^{2}$ in the expression of Eq. (\ref{entasymp}) is made
\textit{as a result of our experimental measurement}\textquotedblright. This
is because the channel selection in Eq. (\ref{X3}) or (\ref{X4})\ cannot be
controlled anyway and happens with the same probability.

A series of paradoxes relevant to quantum entanglement such as the so-called
nonlocality\cite{einstein1935can} follows from the fact that the
Schr\"{o}dinger equation does not have a mechanism to decouple the
entanglement. (Conversely, there is no internal mechanism either for
asymptotically non-entangled states to get entangled according to the
Schr\"{o}dinger equation alone.) A widely accepted postulate to interpret the
nonlocality
\cite{auletta2001foundations,home2007einstein,whitaker2012new,freire2022oxford}
is that upon a physical measurement of one of the two possible asymptotic
functions $a(q_{1}^{\alpha})b(q_{2}^{\beta})$ and $b(q_{1}^{\alpha}%
)a(q_{2}^{\beta})$ the total Schr\"{o}dinger function collapses
instantaneously at both far remote ends $q^{\text{A}}$ and $q^{\text{B}}$
simultaneously. Another postulate is that the action of observation itself
determines what happens in reality. (Einstein allegedly asked
\textquotedblleft do you think is the moon there when nobody
looks?\textquotedblright\cite{mermin1985moon}) By contrast, the quantum
stochastic path approach tells that the channel is selected anyway without
respect to whether the experimental measurement is made or not, and instead
that the intrinsic quantum stochasticity does it spontaneously. [We do not
intend to step into the discussions about Bell's
inequality\cite{auletta2001foundations,home2007einstein,whitaker2012new,freire2022oxford}
in this paper.]

Before concluding this subsection, we raise another elementary example in
which the quantum stochasticity is critical in spontaneous symmetry breaking
in the path solutions. Suppose a dissociation of hydrogen molecule cation
H$_{2}^{+}$ in the following scheme%

\begin{equation}
\left(  \text{H}_{\text{A}}-\text{H}_{\text{B}}\right)  ^{+}\rightarrow%
\begin{array}
[c]{c}%
\nearrow\text{H}_{\text{A}}^{+}+\text{H}_{\text{B}}\\
\\
\searrow\text{H}_{\text{A}}+\text{H}_{\text{B}}^{+}%
\end{array}
\label{H2plus}%
\end{equation}
The Schr\"{o}dinger equation by itself has no mechanism to determine the
branching, which obviously needs to break the spatial symmetry (we here
disregard the nuclear spin). Otherwise the dissociation is forced to
H$_{\text{A}}^{\frac{1}{2}+}$ + H$_{\text{B}}^{\frac{1}{2}+}$ in the density
$\rho$ (only one electron is involved here). There is no inconvenience in
neglecting the physical process behind Eq. (\ref{H2plus}) in the class rooms
for the nature of chemical bond. The view from quantum stochastic path
dynamics accounts for such a symmetry-breaking dissociation in a natural manner.

\color{black}

\section{Summary and concluding remarks \label{sec:Conc}}

We have studied the dual structure of the Schr\"{o}dinger dynamics; one
responsible for the dynamics of quantum distribution amplitude function
subjected to the Schr\"{o}dinger equation and the other for quantum stochastic
path dynamics. To derive the Schr\"{o}dinger equation from scratch on the
field of real number, where the Schr\"{o}dinger function is defined as a
factorizing vector (the Schr\"{o}dinger vector) of a density distribution
function, we have imposed just the minimal requirements on the density;
space-time translational invariance and the flux conservation applied to the
density functions. Then the variational principle applied to the equation of
continuity is shown that the Schr\"{o}dinger equation works to pick the most
likely states among those functions whose densities satisfy the equation of
continuity.\ It is noted that the way of derivation is universal and is not
restricted to quantum mechanics, except for the choice of the Planck constant
and the potential function to build up the concrete form of the Hamiltonian.

We then have shown that the real-valued Schr\"{o}dinger equation gives a
real-valued form of the path integration as its Green function, which is
similar to the Feynman-Kac formula, and thereby highlighting the similarity
between quantum and stochastic dynamics more closely than ever. In contrast to
the Feynman path integrals, it has the Wiener measure and secures the
convergence with respect to the number of sampling paths. The present path
integration supports an anticipation that quantum mechanics bears stochastic
paths behind the Schr\"{o}dinger distribution amplitude function.

The quantum stochastic path dynamics is represented with the Ito stochastic
differential equation consisting of the velocity drift term and the Wiener
process having an appropriate diffusion constant. The velocity drift term
drives the \textquotedblleft particles\textquotedblright\ in a mechanical way,
the resultant pathways of which should geometrically coincide with the Bohmian
trajectory, if the background Schr\"{o}dinger function applied is common, if
the initial conditions are the same, and if there is no Wiener process applied
to the quantum stochastic paths. The Bohm representation of the complex
Schr\"{o}dinger function corresponds to the polar representation of the
Schr\"{o}dinger vector.

The concepts and phenomena derived from quantum stochastic path dynamics are
summarized as follows. (1) Since the local velocity distribution is found to
be a functional of the Schr\"{o}dinger function, indirect correlations arise
among the \textquotedblleft independent\textquotedblright\ quantum paths. The
indirect correlation should therefore depend on the context of experimental
preparation and the resultant Schr\"{o}dinger function. (2) Because of the
properties of the velocity drift terms, the resultant quantum paths
collectively form a fringe pattern in the double slit experiment, just as the
ensemble of Bohm trajectories do. This is not due to an interference among the
quantum stochastic paths. (3) It is clarified that the quantum stochasticity
serves as one of the foundations of quantum dynamics, exemplifying that the
energy eigenvalues of the hydrogen atom and the Bohr radius are given by the
scaling law of the quantum stochasticity alone, without using the Bohr model
or the Schr\"{o}dinger equation. (4) Uncertainty relations of
position-momentum and time-kinetic energy arise as the results of quantum
stochasticity. (5) The quantum Wiener process can spontaneously break the
quantum entanglement and associated symmetry.

We have identified the ultimate mystery or wonder of the Schr\"{o}dinger
dynamics; the relationship between the whole (the Schr\"{o}dinger function)
and its parts (quantum stochastic path dynamics). This relationship is
self-referential and thereby nonlinear: each \textquotedblleft
part\textquotedblright\ operates under instructions\ from the
\textquotedblleft whole\textquotedblright\ (through the velocity drift term),
while the \textquotedblleft whole\textquotedblright\ is composed of a set of
\textquotedblleft parts\textquotedblright\ through the path integration in the
extended Feynman-Kac formula.\ However, it remains unclear how those
individual paths can feel the overall velocity distribution function. In an
effort to unravel this ultimate nonlinear relationship, we explored the
quantum stochastic paths in the quantum Hamilton canonical equations of motion
and the quantum Newtonian equation, thus investigating the quantum-classical
correspondence, yet without reaching a definite conclusion. One rather
comfortable idea to accept the relationship between the whole and parts is to
regard the velocity drift term as a field on which each quantum paths are
guided under a given initial condition. 

Some very brief comments on the dual structure of the Schr\"{o}dinger
dynamics: It could be risky to describe every quantum phenomenon solely in
terms of the Schr\"{o}dinger equation. For example, the interpretation in the
double-slit experiment that a particle can pass through two slits
simultaneously should be incorrect. While the Schr\"{o}dinger function, as a
coherent distribution amplitude function, is mathematically allowed for this,
each quantum stochastic path can physically pass through only one. Similarly,
the spots that appear on the measurement board represent the endpoints of
quantum stochastic paths, and there is no physical need of the so-called
instantaneous collapse of the Schr\"{o}dinger function. The Schr\"{o}dinger
function can predict experimental results that are infinitely repeatable (or
at least performable many times). Conversely, the understanding and prediction
of experiments of limited number of performance, or the theoretical simulation
of non-repeatable phenomena, should resort to the quantum stochastic path dynamics.

Finally, the dual structure of the Schr\"{o}dinger dynamics is expected to be
applied to quantum collision theory, atto-second electron dynamics in atoms
and molecules, quantum chaos to be manifesting in the Schr\"{o}dinger function
and quantum paths, quantum technology, and so on. Also, the classic
interpretations of quantum mechanics such as the Copenhagen interpretation and
others that were figured out to avoid the instantaneous collapse of the
Schr\"{o}dinger
function\cite{auletta2001foundations,home2007einstein,whitaker2012new,freire2022oxford}
are needed to be better-scrutinized. Also, we need to shed a new light on the
theories and interpretation that presuppose the quantum entanglement is never
detangled. These are our next subjects following the present work.\ \color{black}

\bigskip

\bigskip

\begin{acknowledgments}
The author thanks Prof. Satoshi Takahashi and Dr. Kota Hanasaki for valuable
discussions and comments. This work has been supported by JSPS KAKENHI, Grant
No. JP20H00373 and JP24K01435.
\end{acknowledgments}

\bigskip

\subsection*{Conflict of Interest}

The authors have no conflicts to disclose.

\bigskip


%

\end{document}